\documentclass[12pt]{article}

\usepackage{scicite}
\usepackage{graphicx}
\usepackage{times}
\usepackage{url}

\newenvironment{sciabstract}{%
\begin{quote} \bf}
{\end{quote}}

\newcommand{\lens}{E325 }
\newcommand{\lensp}{E325}
\newcommand{\be}{\begin{equation}}
\newcommand{\ee}{\end{equation}}
\newcommand{\dee}[0]{\mathrm{d}}

\title{A precise extragalactic test of General Relativity}

\author
{Thomas E. Collett,$^{1\ast}$ Lindsay J. Oldham,$^{2}$ Russell Smith,$^{3}$
Matthew W. Auger,$^{2}$\\ Kyle B. Westfall$^{1,4}$, David Bacon,$^{1}$ Robert C. Nichol,$^{1}$  Karen L. Masters,$^{1,5}$\\ Kazuya Koyama,$^{1}$ and Remco van den Bosch$^{6}$
\\
\normalsize{$^{1}$ Institute of Cosmology and Gravitation, University of Portsmouth,}\\
\normalsize{Burnaby Rd, Portsmouth, PO1 3FX, UK}\\
\normalsize{$^{2}$ Institute of Astronomy, University of Cambridge,}\\
\normalsize{Madingley Rd, Cambridge, CB3 0HA, UK}\\
\normalsize{$^{3}$ Centre for Extragalactic Astronomy, University of Durham}
\normalsize{Durham DH1 3LE, UK}\\
\normalsize{$^{4}$ University of California Observatories - Lick Observatory,}\\
\normalsize{University of California - Santa Cruz, 1156 High St. Santa Cruz, CA 95064, USA.}\\
\normalsize{$^{5}$ Haverford College, Department of Physics and Astronomy, }\\
\normalsize{370 Lancaster Avenue, Haverford, Pennsylvania 19041, USA}\\
\normalsize{$^{6}$ Max Planck Institute for Astronomy, K\"onigstuhl 17, 69117 Heidelberg, Germany}\\
\normalsize{$^\ast$E-mail:thomas.collett@port.ac.uk}
}

\date{}

\begin{document} 


\baselineskip24pt


\maketitle


\begin{sciabstract}
Einstein's theory of gravity,  General Relativity, has been precisely tested on Solar System scales, but the long-range nature of gravity is still poorly constrained. The nearby strong gravitational lens, ESO~325-G004, provides a laboratory to probe the weak-field regime of gravity and measure the spatial curvature generated per unit mass, $\gamma$. 
By reconstructing the observed light profile of the lensed arcs and the observed spatially resolved stellar kinematics with a single self-consistent model, we conclude that $\gamma = 0.97 \pm 0.09$ at 68\% confidence. 
Our result is consistent with the prediction of 1 from General Relativity and provides a strong extragalactic constraint on the weak-field metric of gravity.
\end{sciabstract}

General Relativity (GR) postulates that mass deforms space-time, such that light passing near to a massive object is deflected. If two galaxies are almost perfectly aligned, the deformation of space-time near the centre of the foreground galaxy can be large enough that multiple images of the background galaxy are observed. Such alignments are called strong gravitational lenses. In the case of a spherical foreground lens and a perfect alignment of lens and source, the background galaxy is distorted into an Einstein ring. The radius of this ring, the Einstein radius, is a function of the mass of the lens, the amount of spatial curvature produced per unit mass and a ratio of three angular diameter distances between the observer, lens and source.

Angular diameter distances are calculated from the redshifts (inferred from the wavelength shift of spectral lines due to the expansion of the Universe) of the lens, $z_l$, and source, $z_s$, and the cosmological parameters of our Universe \cite{planckcosmo}. Therefore the combination of a non-lensing measurement of the mass of a strong lensing galaxy and a measurement of the Einstein radius constrains the amount of spatial curvature produced per unit mass and tests if GR is the correct theory of gravity.

In the limit of a weak gravitational field, the metric of space-time is characterised by two potentials \cite{Bertschinger2011}: the Newtonian potential, $\Phi$, and the curvature potential, $\Psi$, such that the comoving distance element is
\be
\dee s^2=a^2(\tau)\left[-(1+2\Phi) \dee \tau^2+(1-2\Psi)g_{ij} \dee x^i \dee x^j\right]\,
\ee
where $\tau$ is the conformal time, $x^i$ are the space-like coordinates, $g_{ij}$ is the three-metric of constant-curvature spaces, $a(\tau)$ is the scale factor of the Universe.

In GR the two potentials are the same, but many alternative gravity models invoked to explain the accelerated expansion of the Universe (such as $f(R)$ gravity \cite{Hu2007})  predict the ratio of the two potentials ($\gamma=\Psi/\Phi$) to be scale-dependent. These alternative models of gravity remove the need for a dark energy to accelerate the expansion of the Universe. Testing the scale dependence of $\gamma$ is therefore a discriminator between GR and these alternative gravities. 

The motion of non-relativistic objects is governed by the Newtonian potential, whilst the motion of relativistic particles is sensitive to both potentials \cite{Bertschinger2011}. Measuring $\gamma$ therefore requires observations of the motions of both relativistic and non-relativistic particles around the same massive object. 

On Solar System scales the GR prediction for $\gamma$ has been verified to high precision. By measuring the travel time of radio photons passing close to the Sun, the Cassini mission found $\gamma=1 + (2.1 \pm 2.3)\times10^{-5}$ \cite{Bertotti2003}. However the extragalactic constraints on GR are much less precise. On scales of 10-100 megaparsec, $\gamma$ has been constrained to just 20\% precision \cite{Simpson2013, Blake2016, Song2011}, by combining weak gravitational lensing and redshift space distortion measurements.
On megaparsec scales, 30\% constraints on $\gamma$ have been achieved using the mass profiles of clusters \cite{Wilcox2015, clashgr}.

On kiloparsec scales strong gravitational lensing, combined with stellar kinematics in the lens, allows a test of the weak field metric of gravity. The kinematics are sensitive only to the Newtonian potential, whilst the lensing is sensitive to the sum of the potentials. 
The deflection angle of light, $\hat{\alpha}({\bf x})$, caused by a galaxy with surface mass density, $\Sigma({\bf x})$, is given by:
\be
\label{eq:deflections}
\hat{\alpha}({\bf x})=\frac{2 G}{c^2}(1+ \gamma) \int {\mathrm{d}}^2{\bf x^\prime}\Sigma({\bf x^\prime})\frac{{\bf x}-{\bf x^\prime}}{|{\bf x}-{\bf x^\prime}|^2},
\ee
where $G$ is the gravitational constant and $c$ is the speed of light. Here $\gamma$ is explicitly assumed to be constant over the length scales relevant for lensing. In this case the lensing and dynamical masses are related by 
\be
M_\mathrm{dyn} = \frac{1+\gamma}{2} M_\mathrm{lensing}^\mathrm{GR} 
\ee
where $M_\mathrm{dyn}$ is the mass derived from the dynamics and $M_\mathrm{lensing}^{\mathrm{GR}}$ is the mass derived from lensing and assuming GR is correct.

Previous studies have combined Einstein radius measurements with stellar velocity dispersions to infer constraints on $\gamma$ \cite{Bolton2006, Schwab2010}. Recent work \cite{Cao2017} used a sample of 80 lenses to infer $\gamma= 0.995 \pm 0.04 \mathrm{(sta)} \pm 0.25 \mathrm{(sys)}$ at 68\% confidence. The systematic uncertainties are much larger than the statistical uncertainties, as the result relies on assumptions about the density profile of the lenses and the orbits of the stars within the lenses. The large distances to the lenses in these samples ($0.1<z_l<1$) make a more detailed study of their kinematics impractical with current instruments.

\begin{figure}
  \centering
    \includegraphics[width=\columnwidth,clip=True, trim={26mm 5mm 26mm 5mm}]{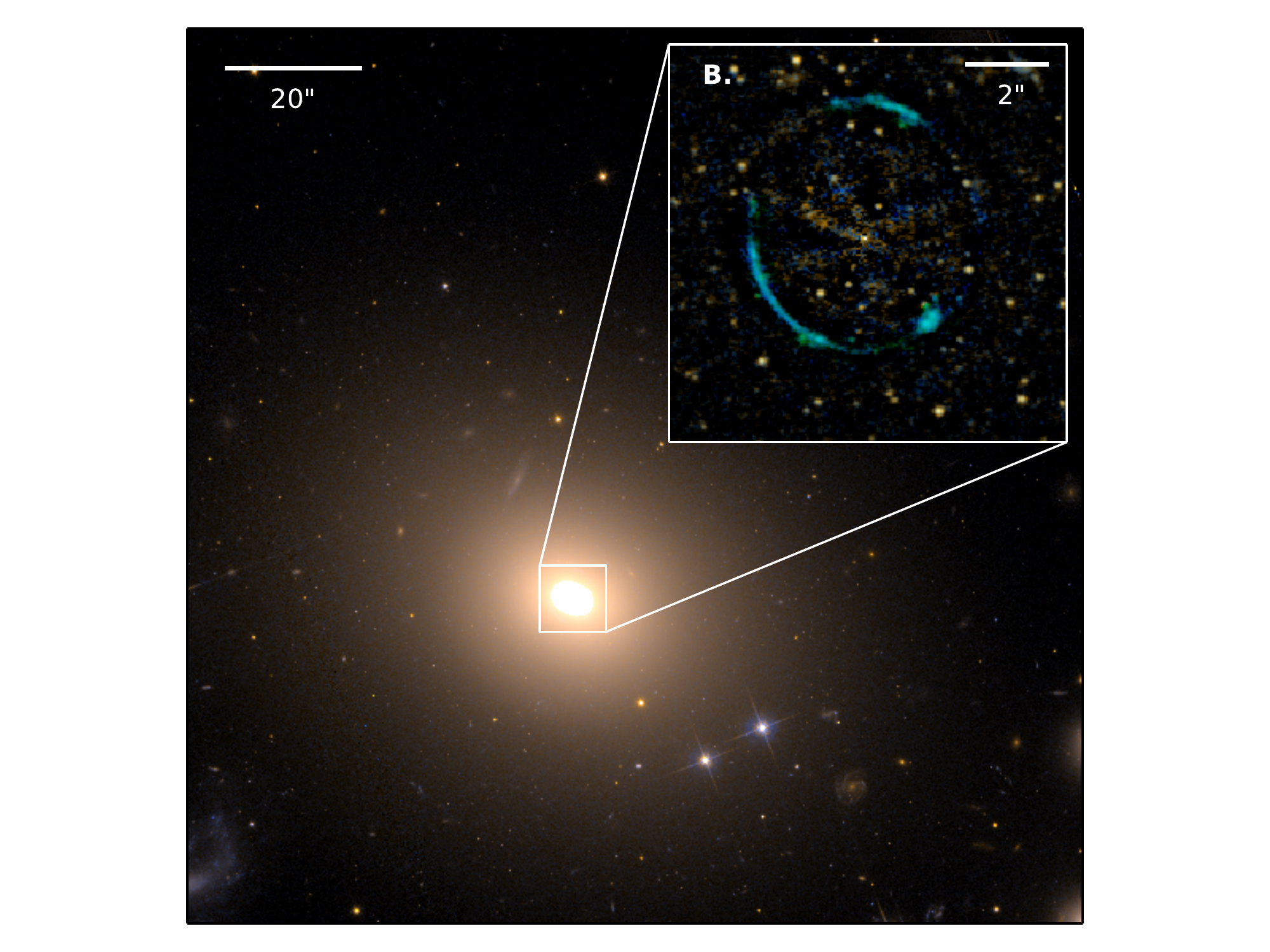}
    \caption {{\bf Colour composite image of ESO325-G004.} Blue, green and red channels are assigned to the F475W, F606W and F814W Hubble Space Telescope imaging. The inset, B, shows a F475W and F814W composite of the arcs of the lensed background source after subtraction of the foreground lens light.}
    \label{fig:colourimage}
\end{figure}

The nearby ($z_l=0.035$) galaxy ESO~325-G004 (hereafter \lensp) is located at right ascension 13:43:33.2, declination $-$38:10:34 (J2000 equinox). Hubble Space Telescope (HST) imaging of \lens serendipitously revealed the presence of an Einstein ring with 2.95 arcsecond radius around the centre of the galaxy \cite{Smith2005},  shown in Figure \ref{fig:colourimage}. Follow-up observations have revealed that the source is an extended star-forming galaxy at $z_s =2.1$ \cite{Smith2013}. \lens provides a laboratory for testing GR on kiloparsec scales as the small distance to the lens enables spatially resolved measurements of the stellar kinematics; this places tight constraints on the 3-dimensional mass structure of the lens. \lens also has an extended arc system, which provides tight constraints on the 2-dimensional surface mass profile of the lens. The small distance also means that the Einstein ring appears in the baryon dominated central region of the lens galaxy: the uncertain distribution of dark matter is much less important than for more distant lenses. 

\lens has previously been observed with the Advanced Camera for Surveys on HST \cite{Smith2005}. We used the HST image observed in the red F814W filter to model the light profile of the galaxy for our mass modelling, and produced an image of the arc light by subtracting the F814W image from that in the blue F475W filter, with the F814W image radially rescaled to account for a slight colour gradient in the lens galaxy \cite{si}. Previous strong lensing constraints on $\gamma$ have relied only upon Einstein radius measurements \cite{Bolton2006,Schwab2010,Cao2017}, yielding a constraint only on the total lensing mass within the Einstein ring. However the extended arcs in \lens also allow for a detailed reconstruction of the unlensed source from the HST observations. This reconstruction places additional constraints on the radial magnification across the image plane wherever lensed images are present \cite{Collettthesis}.  The radial magnification is a weighted integral of the mass within the Einstein ring hence reconstructing the arcs excludes many density profiles that would have the same Einstein radius, but produce arcs of different shapes and widths. This approach is now mature in strong lens studies where high precision constraints are required \cite{Suyu2013, Vegetti2010, Collett2014, Wong2017}.

To constrain the dynamical mass we used observations with the Multi Unit Spectroscopic Explorer (MUSE) \cite{Bacon2010} on the European Southern Observatory (ESO) Very Large Telescope (VLT) to obtain spatially resolved spectroscopy across the lens \cite{si}. From the spectra we extracted the velocity dispersion of the stars in each pixel. We use Jeans axisymmetric modelling \cite{Cappellari2008} to infer the dynamical mass of the lens from this data \cite{si}.

We simultaneously fit a 20 parameter model to both the MUSE and HST data \cite{si}. The parameters describe the mass-to-light ratio of the stellar component observed with HST, including a mass-to-light gradient, a dark matter halo, and a central supermassive black hole. We also include parameters that describe external lensing shear from masses close to the line-of-sight, the radial profile of the stellar orbital anisotropy and the unknown inclination of the lens galaxy with respect to the line-of-sight. The final parameter of our model is $\gamma$, the ratio of the Newtonian and curvature potentials. We assume that this ratio is constant across the relevant length scales of this lens. The smallest scale we resolve corresponds to 100 parsecs in the lens and the Einstein radius corresponds to 2000 parsecs.

\begin{figure}
  \centering
    \includegraphics[width=\textwidth,clip=True]{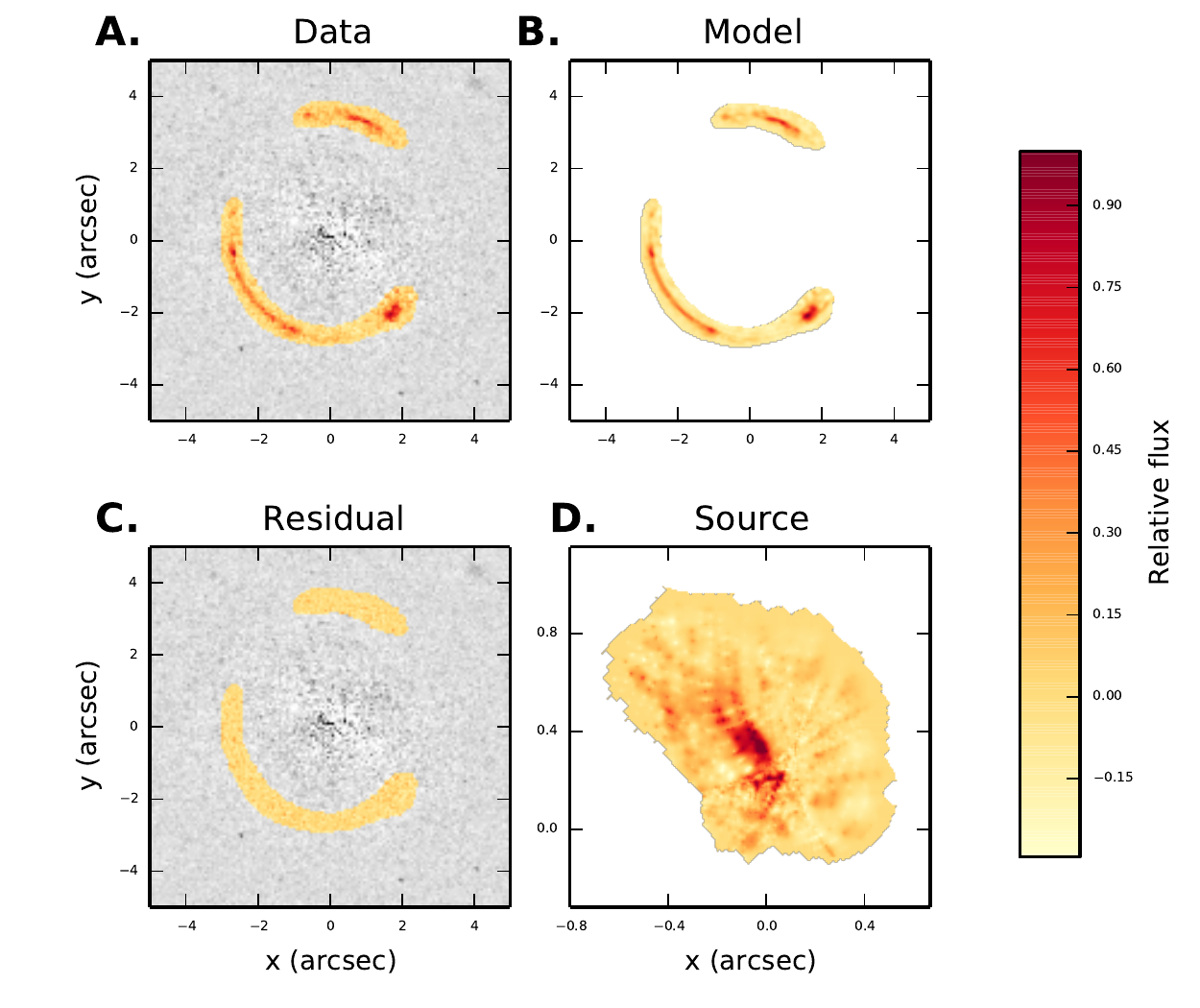}
    \caption{ {\bf Best fitting reconstruction of the lensed arcs in \lensp}. {A.} the foreground subtracted F475W HST image \cite{si}; we analyse only the data shown in colour. {B.} the best fit model of the lens. { C.} the difference after subtracting the model from the data. { D.} the reconstruction of the unlensed source for the best fitting model. In all panels the units are in arcseconds, with the lens centred at (0, 0). The colour bar shows the relative flux for all the images \vspace{1cm}}
    \label{fig:HSTfit}
\end{figure}

\begin{figure}
  \centering
    \includegraphics[width=\textwidth,clip=True]{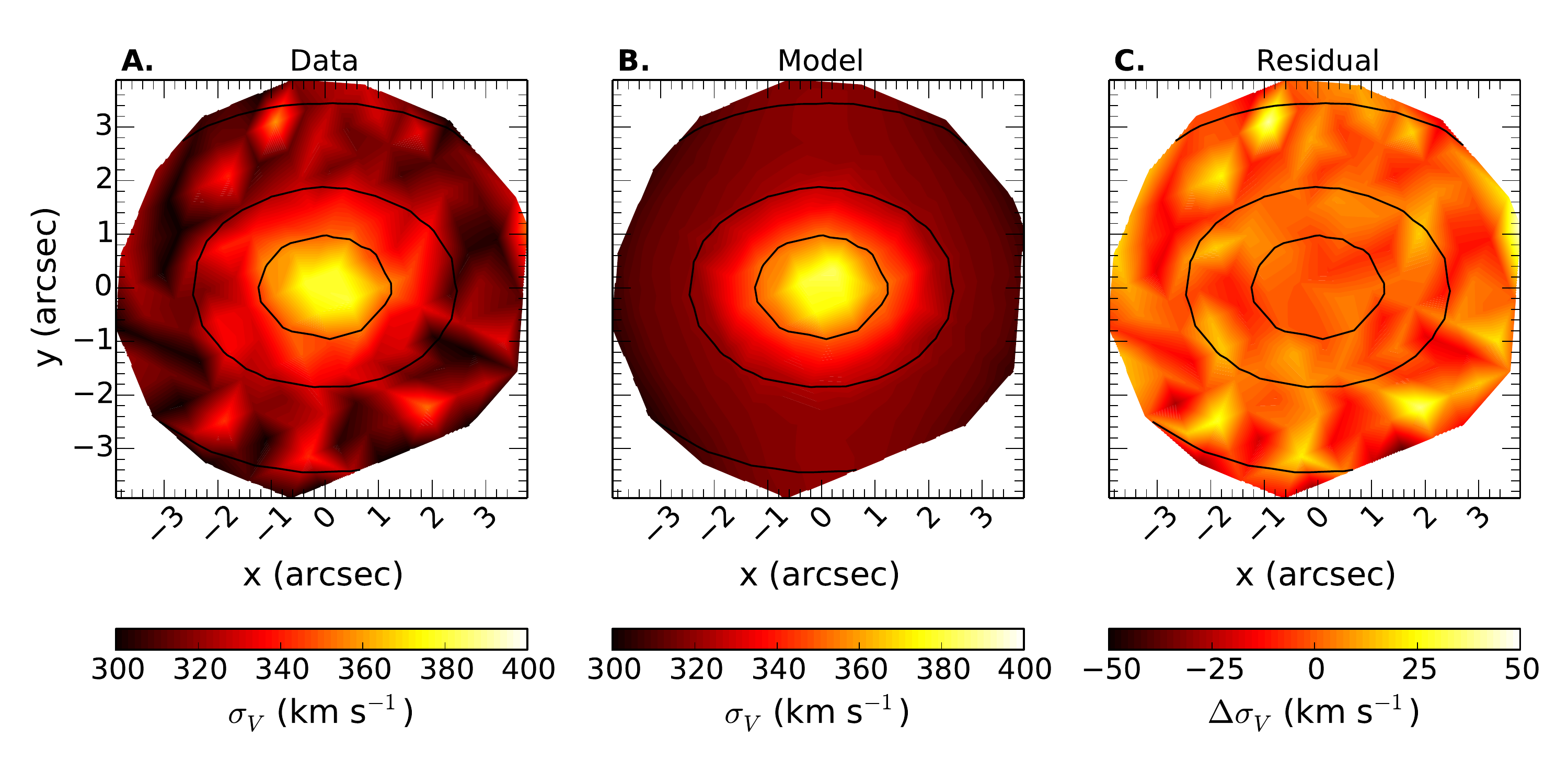}
    \caption {{\bf Two dimensional velocity dispersion profile for \lensp.} {A.} The observed MUSE velocity dispersion data. {B.} Kinematics predicted by our best fitting model. C: The residual difference between the data and model.}
    \label{fig:MUSEfit2d}
\end{figure}

\begin{figure}
  \centering
    \includegraphics[width=0.6\columnwidth,clip=True]{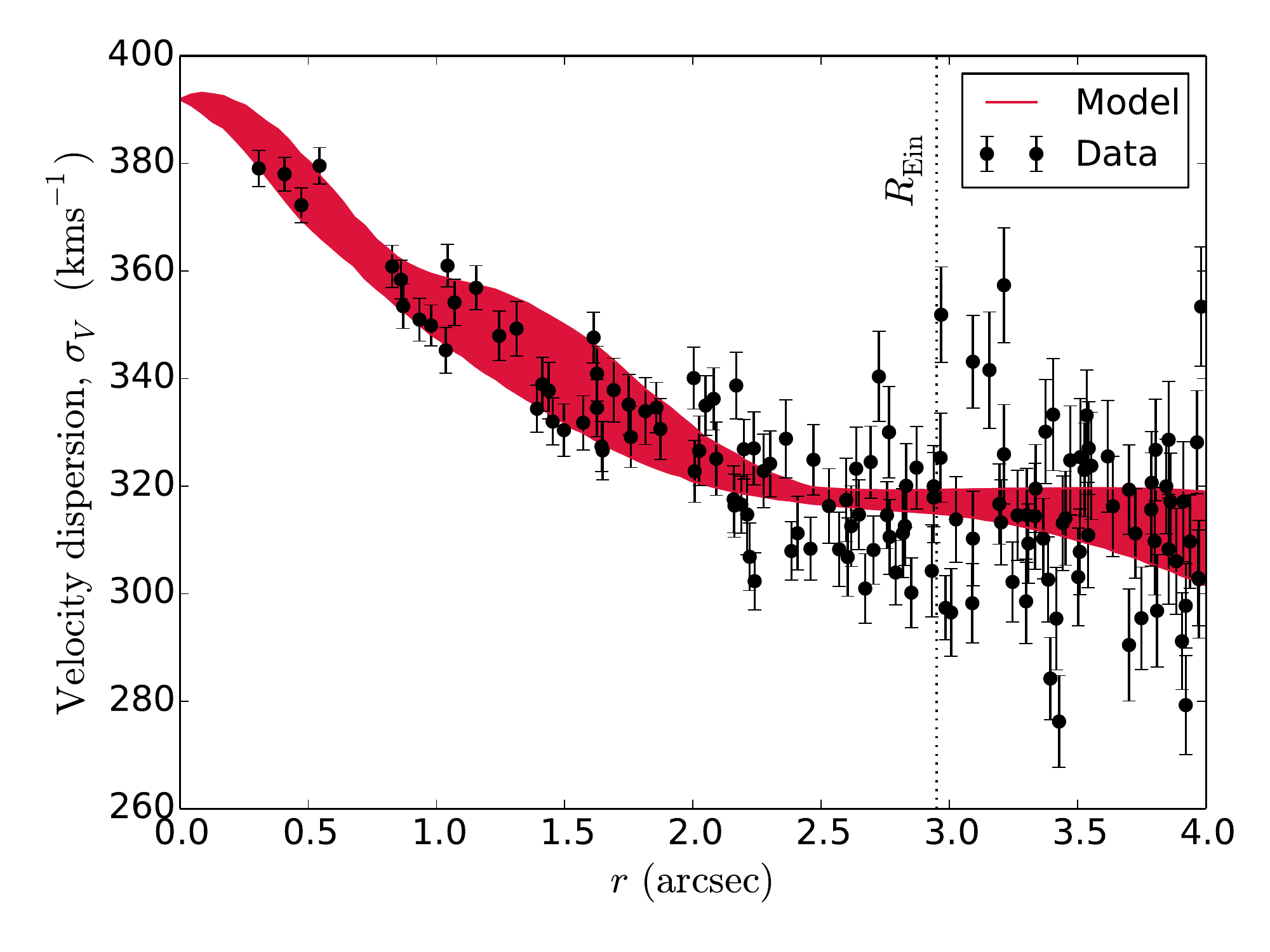}
    \caption {{\bf One dimensional velocity dispersion profile for \lensp, as a function of radial distance from the lens centre}. Black circles show observed values from the MUSE data for the 0.6 arcsecond pixels used in the analysis; error bars are 1 sigma statistical uncertainties. The red band shows the range of velocity dispersions at each $r$ predicted by our best fitting model. The width of the band is due to azimuthal variation in the velocity dispersion, not the statistical uncertainty. The vertical line indicates the Einstein radius.}
    \label{fig:MUSEfit1d}
\end{figure}

By sampling the posterior probability of our model fit to the data, using a Markov Chain Monte Carlo method \cite{emcee}, we infer the uncertainties on the model parameters. Our best fitting reconstruction of the HST image is shown in Figure \ref{fig:HSTfit}. Our model simultaneously reproduces both the observed lensing and dynamical data. Our reconstruction of the MUSE data is shown in Figures \ref{fig:MUSEfit2d} and \ref{fig:MUSEfit1d}. 

Within the constraints of our fiducial model, we infer that it is only possible to simultaneously reconstruct the lensing and kinematics if the stellar mass-to-light ratio of the lens increases within the central kiloparsec, with an observed F814W-band stellar mass-to-light ratio at large radius of $2.8 \pm 0.1 M_\odot/L_{\odot}$ in units of solar mass over solar luminosity in this band (consistent with a Milky Way-like stellar initial mass function [IMF]), rising to $6.6 \pm 0.1 M_\odot/L_{\odot}$ in the centre (consistent with an excess of mass in low luminosity stars relative to the Milky Way). We infer that the central black hole has a mass of $3.8 \pm 0.1 \times 10^{9} M_\odot$, consistent with expectations \cite{McConnell2013} given the mass of \lens . We find that the lens must be inclined close to the plane of the sky, with an inclination angle of $90 \pm 15$ degrees. The typical orbits of the stars are mildly radial at most radii, although they are poorly constrained in the central 0.5 kpc. The dark matter in our fiducial model only accounts for $3^{+3}_{-2}$ percent of the mass within the Einstein radius, consistent with expectations extrapolating from higher redshift lenses \cite{Auger2010b} given the 2.95" Einstein radius is a quarter of the 13" radius that contains half of the light of the lens. 

Our fiducial model assumes that the dark matter follows the Navarro-Frenk-White (NFW) profile \cite{nfw} that is found in cosmological dark-matter only simulations ; however, the baryonic processes of galaxy formation are capable of resculpting the dark matter profile \cite{Blumenthal1986,Gnedin2004,Abadi2010}. An alternative model without a stellar mass-to-light gradient is also able to reconstruct the data, although the central dark matter density would have to fall with radius as $r^{-2.6}$  which is much steeper than the fiducial $r^{-1}$ profile. In this model the baryons are consistent with a Milky Way-like stellar initial mass function and the dark matter accounts for $31 \pm 4 \%$ of the mass within the Einstein radius  \cite{si}. 

We also investigated another model that does not partition the mass into baryonic and dark matter: this model also required a steeply rising total mass-to-light ratio in the central regions. Our current data cannot distinguish between highly concentrated dark matter, a steep stellar mass-to-light gradient or an intermediate solution, but \lens is definitely not consistent with an NFW dark matter halo and constant stellar mass-to-light ratio. 

In our fiducial model, we find $\gamma = 0.978^{+0.010}-{0.015}$ at 68 \% confidence, with the uncertainties estimated from a Markov Chain Monte Carlo analysis. Due to the small statistical uncertainties, we also carry out a thorough analysis of our systematic uncertainty relating to the parameterisation of the lens model.

In principle the parameterisation of the mass profile should not impact the inference on $\gamma$, because both the lensing and kinematics are sensitive to the total mass distribution, rather than the partition between dark matter, black hole and baryons. The alternative models investigated \cite{si} show an excess scatter on $\gamma$ of 0.01, in addition to the statistical uncertainties.

Our model assumes the fiducial cosmology from the Planck satellite's cosmic microwave background observations \cite{planckcosmo}. Our inference on the lensing mass is inversely proportional to the assumed value of the current expansion rate of the Universe, $H_0$. The low redshift of the lens, causes our inference on $\gamma$ to be almost independent of the other cosmological parameters. The 1.3\% uncertainty on the Planck determination of $H_0$ is equivalent to a  2.6\% uncertainty on $\gamma$.

Our dynamical model assumes a particular stellar spectral library to extract the kinematics. For our data there is a 2.2\% systematic shift between velocity dispersions measured with theoretically motivated stellar libraries \cite{CvD12} and those empirically derived from observations of Milky Way stars \cite{INDOUS}. This translates to a 4.4\% uncertainty on the dynamical mass and hence 8.8\% $\gamma$. This systematic dominates the final uncertainty on $\gamma$.

By simultaneously reconstructing the extended light profile of the arcs and spatially resolved kinematics of \lens we precisely constrained $\gamma$ with a single system. The high resolution data allows us to fit for both the density profile and anisotropy profile of the lens, removing these systematics which limited previous strong lensing constraints on $\gamma$ \cite{Bolton2006, Schwab2010, Cao2017}. We conclude that  $\gamma = 0.97 \pm 0.09$ (68 \% confidence limits) including our identified systematics. This constraint on the ratio of the two potentials outside the Solar System confirms the prediction of GR in the local Universe for galactic masses and kiloparsec scales. Our result implies that significant deviations from $\gamma=1$ can only occur on scales greater than $\sim$2 kiloparsecs, thereby excluding alternative gravity models that produce the observed accelerated expansion of the Universe but predict $\gamma \neq 1$ on galactic scales.

\newpage

\section*{Acknowledgements}
\noindent  
We are grateful to Stefano Carniari for help reducing the MUSE data, and to Drew Newman for discussions.
We thank the three anonymous referees for their constructive feedback.

\section*{Funding}
TEC is funded by a Dennis Sciama Fellowship from the University of Portsmouth.
RJS acknowledges support from the STFC through grant ST/P000541/1.
DB, BN, KM and KK are supported by STFC through grant  ST/N000688/1.
KK is supported by the European Research Council (ERC) (grant agreement 646702 "CosTesGrav").
Numerical computations were done on the SCIAMA High Performance Compute (HPC) cluster which is supported by 
the ICG, SEPNet and the University of Portsmouth.

\section*{Author contributions}
\noindent  
TEC performed the lensing and dynamical modelling, lead the proposal for the MUSE data, and wrote the paper. LJO implemented the JAM modelling of \lensp, extracted the MUSE velocity dispersions and co-wrote the paper. RJS performed the initial kinematic fits and produced the lens-subtracted HST data. MWA reduced the HST and MUSE data. KBW helped design and implement the MUSE observations. All authors provided ideas throughout the project and comments on the manuscript and observing proposal.

\section*{Competing interests}
\noindent
The authors have no competing interests.

\section*{Data and materials availability}
\noindent  
This work is based on observations made with
ESO telescopes at the La Silla Paranal Observatory under program ID 097.A-0987(A). This data is available at the ESO archive, \url{archive.eso.org}.
This work is based on observations made with the NASA/ESA Hubble Space Telescope (GO 10429, PI: Blakeslee and GO 10710, PI: Noll), and obtained from the Hubble Legacy Archive, \url{hla.stsci.edu}, which is a collaboration between the Space Telescope Science Institute (STScI/NASA), the Space Telescope European Coordinating Facility (ST-ECF/ESA) and the Canadian Astronomy Data Centre (CADC/NRC/CSA). 
Our lens modelling code {\sc pylens} is available at \url{github.com/tcollett/pylens}. The {\sc mge\_fit\_sectors} software and JAM dynamical modelling code is available at \url{www-astro.physics.ox.ac.uk/~mxc/software/}. The ensemble sampler {\sc emcee} is available at  \url{dfm.io/emcee/}.



\newpage

\def\mnras{Mon. Not. R. Astron. Soc.} 
\def\apj{Astrophys. J. } 
\def\aap{Astron. Astrophys.} 
\def\aj{Astron. J.} 
\def\apjl{Astrophys. J. } 
\def\apjs{Astrophys. J. Suppl. Ser.}
\def\apss{Astrophys. Space Sci.}
\def\aaps{Astron. Astrophys. Suppl. Ser.}
\def\araa{Annu. Rev. Astron. Astrophys.}
\def\nat{Nature} 
\def\prd{Phys.~Rev.~D}
\newcommand{\pasp}{Publ. Astron. Soc. Pac.}
\bibliographystyle{Science}


\newpage

\newpage 


\newpage
\renewcommand\thetable{S\arabic{table}}
\renewcommand\theequation{S\arabic{equation}}
\setcounter{equation}{0}    

\section*{Supplementary Text}
\section{MUSE data reduction and kinematics}
\label{sec:muse}

\lens was observed with MUSE operating in wide-field mode for one hour on 4 April 2016. Five on-source exposures and two sky exposures each of 330 seconds were obtained, and combined using the standard \textsc{esorex} packages \cite{esorex} to extract spectra in the wavelength range $4750-9350\textrm{\AA}$ with 0.2$''$ spatial pixels. 
We fit a Gaussian profile to a nearby unsaturated star to determine the full width at half maximum of the point-spread function (PSF), which was $0.57''$. 

Maps of the first and second order velocity moments were obtained following standard methods \cite{Oldham2017b}. To obtain sufficient signal-to-noise for reliable kinematic measurements, the spectra were binned to $0.6''$ pixels. For each pixel, the wavelength region of $4750-5600\textrm{\AA}$ was modelled as the linear sum of a set of G- and K-type stellar spectral templates from the Indo-US stellar library \cite{INDOUS} -- corresponding to the stellar populations of the lens galaxy. The spectral model also included an additive order-6 polynomial, accounting for residual difference  in the continuum (and any flux from the source galaxy, though this is small), the inferred kinematics are robust to increasing the order of the polynomial.

We quantified the sensitivity of the inferred kinematics to the choice of stellar templates by repeating the inference using the Conroy and van Dokkum (CvD) stellar templates \cite{CvD12} and the Medium resolution INT Library of Empirical Spectra (MILES) templates \cite{MILES}. We found that, whilst the MILES and Indo-US templates give consistent inference, the CvD templates give velocity dispersions which are systematically smaller by $2.2\%$. This reflects the underlying uncertainty in the intrinsic spectra of the stellar populations (see also \cite{Newman2017}). We also verified that our kinematic extraction was robust against variations in the wavelength range being fitted, the masking of features such as the Mg I Fraunhofer b-line (which is not always well reproduced by empirical libraries, \cite{Greene2006}) and continuum parameterisations. We found that each of these changes altered the inferred velocity dispersions by less than 1\%.

There is no evidence for rotation in this system, and  the velocity dispersion increases towards the centre ($\sigma = 371 \pm 3$ kms$^{-1}$ in the central bin).

\section{HST data: reduction and separation of the lens and source light profiles}
\label{sec:HST}
The HST imaging of \lens was obtained with the Advanced Camera for Surveys in 2005 to 2006. We used archival observations: the 18900 sec observation in F814W \cite{GO10429} and the 4800 sec observation in F475W \cite{GO10710}, which is deeper than the original F475W imaging \cite{Smith2005}. The  individual images were processed using the standard ACS pipeline, and combined using the drizzle algorithm \cite{Fruchter2002} as implemented in {\sc MultiDrizzle} \cite{drizzle}.

We exploit the colour contrast between the arcs and the lens to isolate the light from the gravitationally-lensed galaxy. We construct an image of the arcs based on the F475W image after subtracting a scaled version of the deep (18900 sec) F814W image. The scaling incorporates a correction for the radially-averaged median ratio of F475W/F814W, in order to track the radial colour gradient in the lens galaxy.

In order to parameterise the light profile of the lens, we perform a multi-Gaussian expansion (MGE) fit  \cite{Emsellem1994} to the F814W image with the {\sc mge\_fit\_sectors} algorithm \cite{Cappellari2002}. This fit excludes the lensed arcs from the model by interpolating across them, and also accounts for the PSF, giving a model for the true light profile of the lens. The MGE fit for \lens results in 7 Gaussians as described in Table \ref{table:MGEfit}. The 7 MGE components are concentric and co-aligned at an angle of 67.7 degrees east of north.

\begin{table}
\caption{The Multi Gaussian Expansion components of the F814W image of \lensp.} 
\label{table:MGEfit}
\centering 
\begin{tabular}{l c c c c c c c} 
\hline  
MGE component & 1 & 2 & 3 & 4 & 5 & 6 & 7 \\
\hline
relative flux &  1 & 8.39 & 75.9 & 111 & 253 & 484 & 1690 \\
scale (arcsec) & 0.03 & 0.19& 0.42 & 0.80 & 1.29 & 2.60 & 7.00 \\
axis ratio & 1. & 0.90  & 0.86 & 0.68 & 0.76 & 0.73 & 0.86 \\
\hline  
\end{tabular}\\

\end{table}

\section{Simultaneously modelling the lensing and dynamical data}
\label{sec:massmodel}

We simultaneously model the HST imaging of the lensed arcs and the MUSE kinematics within the Einstein radius with a single self-consistent model for the mass distribution of the lens. We assume that the mass is composed of a luminous baryonic component, a dark matter component and a central black hole.

We assume that the baryonic component follows the observed luminosity distribution of the lens, modulated by a varying stellar-mass-to-light ratio. We use the MGE fit described in Section \ref{sec:HST} to decompose the light profile into a model for which lensing and dynamical quantities can both be computed efficiently \cite{Barnabe2012,Cappellari2008}. Each component of the MGE is given a parameter to describe its mass-to-light ratio, $\Upsilon_i$, and these parameters are constrained to decrease with the radius of the MGE component; this ensures that the total baryonic mass-to-light profile is monotonically decreasing. Since the two central MGEs have radii smaller than the MUSE PSF, we fix their mass-to-light ratios to be the same. Our baryonic model thus has 6 parameters.

We assume that the dark matter is concentric with the baryonic matter and has the same alignment. We model its density profile using the pseudo-NFW profile \cite{Munoz2001}:

\be
\rho(r) = \frac{\rho_0}{r (r_s^2+r^2)}.
\label{eq:pnfwprofile}
\ee
where $\rho(r)$ is the density as a function of three-dimensional distance from the centre, $\rho_0$ is the central density, and $r_s$ is the scale radius of the halo. We use a pseudo-NFW profile with $(r_s^2+r^2)$ in the denominator rather than $(r_s + r)^2$ as this has the computational advantage that the deflection angles can be calculated analytically.

For computational efficiency we include ellipticity in the lensing potential rather than the density profile; this is a good approximation for almost spherical haloes \cite{Barkana1998}. We make the approximation that the flattening of the potential is half the flattening of the 2D surface mass density, such that $q_\mathrm{potential}=1-(1-q_\mathrm{density})/2$ where $q$ is the axis ratio \cite{Barkana1998}. Our results do not change if we make the much worse approximation $q_\mathrm{potential}=q_\mathrm{density}$. 

Since we are only modelling the kinematics within the central 4 arcseconds of the lens, the much larger scale radius of the dark matter halo is not constrained by our data: we therefore fix it to 10 times the half light radius of the lens, as is typical of haloes in dark matter simulations \cite{Kravtsov2013}. Our dark matter component therefore has two free parameters, the axis ratio of the halo ($q_\mathrm{DM}$) and the amplitude of the dark matter profile, which we quantify in terms of the fraction of mass within the Einstein radius which is provided by dark matter ($f_\mathrm{DM}$).

We model the black hole as a point mass with mass $M_\mathrm{BH}$ at the centre of the lens as inferred from the MGE fit in Section \ref{sec:HST}.

In addition to the mass parameters, the model has 10 further free parameters. Two parameters describe the lensing effect of structures near to the line-of-sight: an external shear magnitude, $\gamma_\mathrm{ext}$, and angle, $\theta_\mathrm{ext}$. The other parameters are relevant for the kinematics: one for the inclination of the lens with respect to the line of sight, $i$, and seven describe the orbital anisotropy profile of the stars, $\beta_i$ (see Section \ref{sec:kinematics}).

The final parameter of our model is $\gamma$. We assume that $\gamma$ is constant across the relevant scales of \lensp.  Given a mass distribution, the deflection angles are derived from Equation \ref{eq:deflections}.

\subsection{Lens modelling}

In order to fully exploit the information content of the lensed arcs, we reconstruct the observed light profiles of the lensed images.
This approach is computationally expensive, but allows us to use the observed fluxes in thousands of pixels as constraints on the lens model. By reconstructing all of these pixels we are able to greatly reduce the range of mass models that fit the data compared to a single Einstein radius estimate. Only a small subset of physically plausible mass models are able to reproduce the observed width and curvature of the arcs and the Einstein radius. This results in a precise constraint on the 2 dimensional surface mass density profile of the lens.

It is clear from the multiple surface brightness peaks in the HST imaging that the source must have multiple clumps. We therefore use a pixellated grid to describe the unlensed source, rather than a simple parametric model. 
Our modelling adopts a semi-linear approach
\cite{Warren2003}, where for each iteration of the mass model, we linearly solve for the optimal
pixellated source to calculate a likelihood of the data given the mass model.

We use an adaptive grid of 80 by 80 square pixels\cite{Collett2014}. Since the source crosses a caustic curve, regions of high magnification are present hence a high resolution source is needed to reproduce substructures in the observed arcs. We ensure the source is smooth using quadratic gradient regularisation \cite{Suyu2006}; this ensures that rapid spatial variations are penalised. We use a Bayesian prescription \cite{Suyu2006} to infer the optimal amount of penalisation directly from the data. 

\subsection{Kinematic Modelling}
\label{sec:kinematics}

We model the stellar kinematics by solving the axisymmetric Jeans equation using the Jeans Anisotropic Models of stellar kinematics or proper motions of axisymmetric or spherical galaxies
 (JAM) implementation \cite{Cappellari2008}. Our aim is to compare lensing and dynamical masses, we therefore model only the central 4 arcseconds of kinematic data. This reduces our ability to break degeneracies between dark and baryonic mass, but it removes the possibility that model miss-match in the dynamics far outside the Einstein radius could bias our inference on $\gamma$.

JAM requires that the gravitational potential be parameterised as the sum of gaussian ellipsoids (the MGE approximation). For the NFW haloes, we precompute an MGE fit to $\rho(r)$ as a function of halo scale radius. At each step in the chain, the overall amplitude of the dark matter MGEs is rescaled to give the correct $f_{DM}$. The dark matter MGEs are then added to the black hole and baryonic components to give the total gravitational potential.



Finally, our combination of strong lensing and kinematics, each of which provides independent and complementary information on the mass structure, allows us to overcome the degeneracy between the gravitational potential and the orbital anisotropy of the tracers. We therefore construct a profile for the anisotropy $\beta(R)$ using the (luminous) MGE, each component of which is allowed an independent value of $\beta$. This is a convenient way to parameterise a varying anisotropy profile but the luminous MGE components should not be interpreted as physically distinct dynamical populations. The anisotropy profile can then be reconstructed from the $\beta_i$ \cite[their Equation 24]{Cappellari2008}, however individual $\beta_i$ are not physically meaningful quantities.

\subsection{Sampling the likelihood function}

Since the lensing and dynamical data are independent, we can multiply the individual likelihoods to produce a single likelihood function. The posterior distributions of the non-linear parameters in our model are probed using a Markov Chain Monte Carlo method. We use the ensemble sampler {\sc emcee} \cite{emcee} to efficiently sample the 20-dimensional parameter space and mitigate against any multi-modality of the posterior. We performed 5,120,000 evaluations of the likelihood function to ensure we fully probed the degeneracies of the parameter space.

\section{Results and Astrophysical Interpretation}

\begin{table}
\caption{The infered median and 68\% confidence intervals for the parameters of our fiducial model.} 
\label{table:1dparams}
\centering 
\begin{tabular}{l l l l l l} 
\hline
\vspace{-3mm}\\
$\Upsilon_\infty$ & $2.8^{+0.1}_{-0.1}$ & Mass-to-light ratio at large radii\vspace{1mm}\\
$\Upsilon_{1,2}$ & $9.5^{+0.5}_{-0.5}$ & Mass-to-light ratio of MGE components 1 and 2 \vspace{1mm}\\
$\Upsilon_3$ & $8.5^{+0.4}_{-0.5}$ &  Mass-to-light ratio of MGE component 3 \vspace{1mm}\\
$\Upsilon_4$ & $3.8^{+0.4}_{-0.3}$ &  Mass-to-light ratio of MGE component 4 \vspace{1mm}\\
$\Upsilon_5$ & $3.4^{+0.1}_{-0.1}$ &  Mass-to-light ratio of MGE component 5 \vspace{1mm}\\
$\Upsilon_6$ & $3.2^{+0.1}_{-0.1}$ &  Mass-to-light ratio of MGE component 6 \\
\vspace{-3mm}\\
\hline
\vspace{-3mm}\\
$\gamma_\mathrm{ext}$ & $0.020^{+0.002}_{-0.003}$ & External shear strength \vspace{1mm}\\\
$\theta_\mathrm{ext}$ & $119^{+2}_{-3}$ & Shear angle, in degrees East of North \\
\vspace{-3mm}\\
\hline
\vspace{-3mm}\\
$M_\mathrm{BH}$ & $3.8^{+0.2}_{-0.2}$ & Black hole mass, in $M_\odot/10^9$\\
\vspace{-3mm}\\
\hline
\vspace{-3mm}\\
$f_\mathrm{DM}$ & $0.033^{+0.034}_{-0.021}$ & Dark matter fraction within $R_\mathrm{Ein}$ \vspace{1mm}\\
$q_\mathrm{DM}$ & $0.74^{+0.19}_{-0.32}$ & Flattening of dark matter halo \\
\vspace{-3mm}\\
\hline
\vspace{-3mm}\\
$i$& $90^{+15}_{-15}$ & Inclination of lens in degrees relative to line-of-sight \\
\vspace{-3mm}\\
\hline
\vspace{-3mm}\\
$\beta_1$ & $-0.6^{+1.2}_{-2.2}$ & Anisotropy parameter for MGE component 1\vspace{1mm}\\
$\beta_2$ & $-1.0^{+1.3}_{-1.8}$ &  Anisotropy parameter for MGE component 2 \vspace{1mm}\\
$\beta_3$ & $0.34^{+0.05}_{-0.08}$ &  Anisotropy parameter for MGE component 3 \vspace{1mm}\\
$\beta_4$ & $-3.4^{+1.4}_{-1.2}$ &  Anisotropy parameter for MGE component 4 \vspace{1mm}\\
$\beta_5$ & $0.39^{+0.02}_{-0.02}$ &  Anisotropy parameter for MGE component 5 \vspace{1mm}\\
$\beta_6$ & $-0.31^{+0.10}_{-0.10}$ &  Anisotropy parameter for MGE component 6 \vspace{1mm}\\
$\beta_7$ & $0.36^{+0.03}_{-0.03}$ &  Anisotropy parameter for MGE component 7\\
\hline
\vspace{-3mm}\\
$\gamma$ & $0.978^{+0.010}_{-0.015}$ & Spatial curvature per unit mass \\
\vspace{-3mm}\\
\hline

\end{tabular}\\
\end{table}

The best fitting parameters of our model and their uncertainties are summarised in Table \ref{table:1dparams}. The two dimensional posteriors for $\gamma$ and parameters that are partially degenerate with it are shown in Figure \ref{fig:corner}.

\lens has an unusual amount of information that is available with which to constrain its mass, which has led to the conclusion that this system has a stellar mass-to-light ratio consistent with expectations assuming a Milky Way-like IMF \cite{Smith2013}. Independent analyses of other galaxies using strong lensing and dynamics have found strong evidence that the stellar mass-to-light ratios of massive early type galaxies require a heavier IMF than the Milky Way \cite{Auger2010b,Cappellari2012}, which stellar population modelling studies have attributed to the IMF being bottom-heavy. There are suggestions that the IMF becomes heavier than Milky Way with galaxy velocity dispersion \cite{LaBarbera2015,MartinNavarro2015,Conroy2012b}, however lensing and dynamical modelling of the three lowest-redshift lenses (including \lensp) has shown that these systems require Milky Way-like IMFs despite their high velocity dispersions. For \lens it has been found that the stellar-mass-to-light ratio is inconsistent with being more bottom-heavy than the Milky Way at the $>4\sigma$ level  \cite{Newman2017}. 

Inference on $\Upsilon_{\star}$ from stellar population modelling of the optical spectrum of \lens over an approximately-matched aperture is inconsistent with that inferred from lensing and dynamics at the $3.4\sigma$ level \cite{Newman2017}. The existence of radial gradients in $\Upsilon_{\star}$ could play an important role in explaining this discrepancy \cite{Newman2017}. Our model attempts to quantify any gradients that may exist.

Firstly, we rule out a spatially uniform stellar mass-to-light ratio at greater than 99\% confidence. As shown in Table 2 and Figure \ref{fig:MLprofile}, we find that $\Upsilon_{\star}$ decreases by a factor of two from $\Upsilon_{\star}(R=0) = 6.6 \pm 0.1$ to $\Upsilon_{\star}(R=R_{Ein}) = 2.8 \pm 0.1$. The luminosity-weighted average stellar mass-to-light ratio within the Einstein radius is $\Upsilon_{\star}(R<R_{Ein}) = 3.8_{-0.2}^{+0.1}$, consistent with previous results \cite{Newman2017,Smith2013} where a spatially uniform $\Upsilon_{\star}$ was assumed. Our inference on the inner stellar-mass-to-light ratio components is not strongly degenerate with either the black hole mass or the anisotropy (Figure \ref{fig:cornerML}). One limitation of our current model is that we have not explored the degeneracy between the variation in stellar mass-to-light ratio and the inner halo slope, which here is fixed to $-1$. In Section \ref{sec:sysmass} we show that no stellar mass-to-light gradient is required if the central dark matter density profile is much steeper than NFW. Constraints on the stellar dynamics at larger radii could  simultaneously constrain  the inner dark matter slope and the stellar mass-to-light profile. However we have not included the larger radius constraints in our modelling, since a model mismatch outside the Einstein radius could potentially bias the inference of $\gamma$.

Variations in $\Upsilon_{\star}$ indicate variations in the properties of the stellar populations as a function of radius; whilst we have not performed the stellar population modelling of the MUSE spectra that could robustly disentangle the contributions from stellar age, metallicity and IMF, we note that the stellar mass-to-light profile $\Upsilon_{\star}(R)$ shown in Figure \ref{fig:MLprofile} is consistent with the IMF of the stars being super-Salpeter in the centre ($\alpha_{\star} = \Upsilon_{\star}/\Upsilon_{MW} = 2.2 \pm 0.1$, based on the stellar population modelling \cite{Smith2013}, which assumed a Kroupa IMF) and roughly Milky Way-like at large radii ($\alpha_{\star}(R=R_{Ein}) = 1.3 \pm 0.1$ and $\alpha_{\star}(R\to\infty) = 0.9 \pm 0.1$). This interpretation would be consistent with previous results based on stellar population modelling \cite{MartinNavarro2015, vdc2017} and dynamics \cite{OldhamM87}; however, even if the IMF is not the driving property behind the inferred trend, such strong variations in any stellar population property imply that the physical conditions during star formation must have varied across the galaxy.

Our flexible mass model also allows us to infer the dark matter fraction and the black hole mass. We find a black hole mass $M_{BH} = (3.8 \pm 0.1) \times 10^{9} M_{\odot}$, consistent with the $M_{BH}-\sigma_{\star}$ relation \cite{McConnell2013}. The projected dark matter fraction within the Einstein radius is $f_{DM} = 0.03_{-0.02}^{+0.03}$, which is low compared to expectations based on the cosmological hydrodynamical simulations \cite{Smith2015}, but consistent with results for other strong lenses \cite{Auger2010a}, given the physically small Einstein radius of \lensp. We find that the stellar orbits are mildly radially anisotropic ($\beta \sim 0.2$ at all radii) with small radial variations, consistent with other ETGs \cite{Cappellari2007}. The inferred anisotropy profile is shown in Figure \ref{fig:beta}. The radial variations of anisotropy shown in Figure \ref{fig:beta} are similar to those seen in high resolution simulations of elliptical galaxies \cite{Wu2014}. Finally we find that the external shear is small, which reflects the fact that \lens is the dominant member of a galaxy group with no group members within 30 arcseconds of the centre of mass.

\section{Systematics}

Our inference on $\gamma$ from the combined lensing and dynamics analysis leads to small statistical uncertainties on $\gamma$. The uncertainty on $\gamma$ is therefore dominated by systematic effects, which we quantify below.

\subsection{Alternative Mass Models}
\label{sec:sysmass}

\begin{table}
\caption{The inferred median and 68\% confidence intervals for the parameters of our first alternative model, fitting a constant stellar mass-to-light ratio and a generalised-NFW dark matter halo.} 
\label{table:1dparams}
\centering 
\begin{tabular}{l l l l l l} 
\hline
\vspace{-3mm}\\
$\Upsilon_\star$ & $3.19^{+0.02}_{-0.06}$ & Stellar Mass-to-light ratio\vspace{1mm}\\
\vspace{-3mm}\\
\hline
\vspace{-3mm}\\
$\gamma_\mathrm{ext}$ & $0.020^{+0.001}_{-0.001}$ & External shear strength \vspace{1mm}\\\
$\theta_\mathrm{ext}$ & $148^{+1}_{-1}$ & Shear angle, in degrees East of North \\
\vspace{-3mm}\\
\hline
\vspace{-3mm}\\
$M_\mathrm{BH}$ & $12.35^{+0.03}_{-0.04}$ & Black hole mass, in $M_\odot/10^9$\\
\vspace{-3mm}\\
\hline
\vspace{-3mm}\\
$f_\mathrm{DM}$ & $0.31^{+0.04}_{-0.04}$ & Dark matter fraction within $R_\mathrm{Ein}$ \vspace{1mm}\\
$q_\mathrm{DM}$ & $0.31^{+0.04}_{-0.04}$ & Axis-ratio of dark matter halo \vspace{1mm}\\
$r_s$ & $1.0^{+0.1}_{-0.1}$ & Scale radius of dark matter halo in kpc \vspace{1mm}\\
$\alpha$ & $2.61^{+0.02}_{-0.02}$ & Central slope of dark matter halo \vspace{1mm}\\
\vspace{-3mm}\\
\hline
\vspace{-3mm}\\
$i$& $90^{+7}_{-7}$ & Inclination of lens in degrees relative to line-of-sight \\
\vspace{-3mm}\\
\hline
\vspace{-3mm}\\
$\beta_1$ & $-0.0^{+0.5}_{-0.4}$ & Anisotropy parameter for MGE component 1\vspace{1mm}\\
$\beta_2$ & $0.91^{+0.01}_{-0.03}$ &  Anisotropy parameter for MGE component 2 \vspace{1mm}\\
$\beta_3$ & $-0.9^{+0.2}_{-0.4}$ &  Anisotropy parameter for MGE component 3 \vspace{1mm}\\
$\beta_4$ & $-1.7^{+0.3}_{-0.2}$ &  Anisotropy parameter for MGE component 4 \vspace{1mm}\\
$\beta_5$ & $0.43^{+0.02}_{-0.02}$ &  Anisotropy parameter for MGE component 5 \vspace{1mm}\\
$\beta_6$ & $-0.35^{+0.09}_{-0.09}$ &  Anisotropy parameter for MGE component 6 \vspace{1mm}\\
$\beta_7$ & $0.32^{+0.02}_{-0.02}$ &  Anisotropy parameter for MGE component 7\\
\hline
\vspace{-3mm}\\
$\gamma$ & $1.007^{+0.013}_{-0.009}$ & Spatial curvature per unit mass \\
\vspace{-3mm}\\
\hline

\end{tabular}\\
\end{table}

\begin{table}
\caption{The infered median and 68\% confidence intervals for the parameters of our second alternative model combining the dark and baryonic matter into a single total density profile.} 
\label{table:1dparams}
\centering 
\begin{tabular}{l l l l l l} 
\hline
\vspace{-3mm}\\
$\Upsilon_\infty$ & $2.89^{+0.06}_{-0.06}$ & Mass-to-light ratio at large radii\vspace{1mm}\\
$\Upsilon_{1,2}$ & $7.9^{+0.5}_{-1.3}$ & Mass-to-light ratio of MGE components 1 and 2 \vspace{1mm}\\
$\Upsilon_3$ & $10.5^{+0.4}_{-0.3}$ &  Mass-to-light ratio of MGE component 3 \vspace{1mm}\\
$\Upsilon_4$ & $4.2^{+0.3}_{-1.9}$ &  Mass-to-light ratio of MGE component 4 \vspace{1mm}\\
$\Upsilon_5$ & $1.1^{+0.6}_{-0.3}$ &  Mass-to-light ratio of MGE component 5 \vspace{1mm}\\
$\Upsilon_6$ & $5.0^{+0.3}_{-0.2}$ &  Mass-to-light ratio of MGE component 6 \\
\vspace{-3mm}\\
\hline
\vspace{-3mm}\\
$\gamma_\mathrm{ext}$ & $0.008^{+0.003}_{-0.003}$ & External shear strength \vspace{1mm}\\\
$\theta_\mathrm{ext}$ & $148^{+2}_{-4}$ & Shear angle, in degrees East of North \\
\vspace{-3mm}\\
\hline
\vspace{-3mm}\\
$M_\mathrm{BH}$ & $3.7^{+0.1}_{-0.3}$ & Black hole mass, in $M_\odot/10^9$\\
\vspace{-3mm}\\
\hline
\vspace{-3mm}\\
$i$& $90^{+10}_{-10}$ & Inclination of lens in degrees relative to line-of-sight \\
\vspace{-3mm}\\
\hline
\vspace{-3mm}\\
$\beta_1$ & $-1.3^{+1.2}_{-2.3}$        & Anisotropy parameter for MGE component 1\vspace{1mm}\\
$\beta_2$ & $-2.3^{+1.8}_{-1.8}$       &  Anisotropy parameter for MGE component 2 \vspace{1mm}\\
$\beta_3$ & $0.35^{+0.06}_{-0.09}$  &  Anisotropy parameter for MGE component 3 \vspace{1mm}\\
$\beta_4$ &  $-3.9^{+1.5}_{-0.7}$       &  Anisotropy parameter for MGE component 4 \vspace{1mm}\\
$\beta_5$ & $0.34^{+0.02}_{-0.03}$  &  Anisotropy parameter for MGE component 5 \vspace{1mm}\\
$\beta_6$ & $-0.15^{+0.09}_{-0.10}$&  Anisotropy parameter for MGE component 6 \vspace{1mm}\\
$\beta_7$ & $0.31^{+0.03}_{-0.05}$  &  Anisotropy parameter for MGE component 7\\
\hline
\vspace{-3mm}\\
$\gamma$ & $0.979^{+0.009}_{-0.016}$ & Spatial curvature per unit mass \\
\vspace{-3mm}\\
\hline

\end{tabular}\\
\end{table}

Our fiducial mass model assumes that the dark matter follows an NFW profile with scale radius ten times the half-light radius $r_s = 10 R_\mathrm{eff}$. However, it is possible that the rigidity of our dark matter mass model could lead to an artificial breaking of the mass-anisotropy degeneracy in the dynamical modelling; furthermore, the fraction of dark matter within the Einstein radius is partially degenerate with $\gamma$ (see Figure \ref{fig:corner}). Whilst our data do not constrain the large scale profile of the dark matter distribution, the choice of dark matter profile may therefore be a systematic for our inference on $\gamma$. 

To assess the importance of these effects, we test our assumptions that the NFW scale radius is large and that the density slope falls as $r^{-1}$.

Fixing the dark matter scale radius to 5 $R_\mathrm{eff}$ yields $\gamma = 0.971 \pm 0.012$. Allowing the scale radius to be a free parameter gives $\gamma = 0.965 \pm 0.013$.

If we fit the dark matter with a generalised NFW profile \cite{Wyithe2001,Munoz2001}, with a free inner slope $\alpha$
\be
\rho(r) = \frac{\rho_0}{r^{\alpha} (r_s^2+r^2)^{(3-\alpha)/2}},
\label{eq:dplprofile}
\ee
and a constant mass-to-light ratio for the baryons, we find that $\gamma$ is $1.007 \pm 0.010$, at 68 percent confidence. Both the free scale-radius and generalised NFW models give very concentrated dark matter distributions: the scale radius of the former is $1.1^{+0.2}_{-0.1}$ kpc, whilst the gNFW model has a scale radius of $1.0 \pm 0.1$ kpc and an inner slope of $2.61 \pm 0.02$. These models also have markedly more dark matter within the Einstein radius: the gNFW model has $f_\mathrm{DM}=0.31\pm0.04$. Such halos would require an extreme amount of dark matter contraction. This amount of contraction is not seen in hydrodynamical simulations: for example Duffy et al \cite{Duffy2010} finds inner dark matter slopes of between -1.5 and -2 for $z=0$ haloes. 

The final model we test is more agnostic to the partition of mass into dark matter and baryons: we fit only the total mass profile by assigning a free mass-to-light ratio for each of the MGE components. We add no dark matter halo and allow any dark matter to be absorbed as part of the MGE. For this model we infer $\gamma = 0.994^{+0.009}_{-0.015}$. Figure \ref{fig:remcoland} shows the total mass-to-light profile inferred for this model. Again we see a sharp increase in the central mass-to-light ratio, indicative of either a stellar mass-to-light gradient or a high concentration of dark matter. Whilst the over-concentrated dark matter halos are unlikely to be correct, an intermediate scenario with somewhat over concentrated dark matter and a mild mass-to-light gradient is also plausible. 

The scatter between these alternative mass models gives us an understanding of the systematic uncertainty on $\gamma$ that is introduced by our model assumptions. The standard deviation between the value of  $\gamma$ inferred for the models is consistent with an additional modelling systematic of 0.01 on $\gamma$.

\subsection{Weak Lensing by the Lens environment and along the line-of-sight}

In addition to mass on the lens plane, we must also account for the effect of lensing by line-of-sight structures. Our model includes the external shear from line-of-sight structures, but neglects external convergence, as this cannot be inferred from lensing data \cite{MSD}. External convergence would alter the Einstein radius of the lens whilst leaving the kinematics unchanged. A full reconstruction of such structures for \lens \cite{Rusu2017,Collett2013} is not possible with existing data, however we argue that this effect is negligible in the case of \lensp, since for a low redshift lens, the effect is suppressed unless the perturber has very similar redshift to the lens. 

Adding a sheet of mass with a small convergence $\kappa_p$ at redshift $z_p$ is equivalent to adding a mass sheet on the lens plane with a smaller amplitude \cite{Keeton2003}:
\be
\kappa_{\mathrm{effective}} = (1-\beta(z_1, z_2)) \kappa_p,
\ee
where $z_1$ and $z_2$ are the smaller and larger of $z_l$ and $z_p$ and 
\be
\beta(z_1, z_2) = \frac{D_{z_1,z_2} D_{0,z_s}}{D_{z_1,z_s}D_{0,z_2}}
\ee
where the $D_{i,j}$ are angular diameter distances between redshifts $i$ and $j$. For the redshifts of \lensp, any perturber below $z=0.02$ or above $z=0.05$ has its effective convergence reduced by at least a factor of two, with perturbers above $z=0.15$ having their effective convergence reduced by at least a factor of 5. Line-of-sight structures are therefore highly unlikely to bias our inference on $\gamma$. This result is consistent with the small external shear inferred from our model: large effective external convergences would require the presence of a cluster very close to the line of sight, which would also generate a large shear. 

\lens is the brightest galaxy in a small group \cite{Smith2005}. We have assumed in our model that the environment of the lens can be described by a shear term; we do not include higher order terms in the deflection angles introduced by group members. This approximation is reasonable since the angular distance to other group members is much greater than the scale of the arcs - there are no group members observed within ten Einstein radii. Low redshift masses are very inefficient lenses since for a much more distant source the critical surface mass density is inversely proportional to angular diameter distance to the lensing mass. Because of this, the mean surface mass density within the Einstein radius of \lens is much higher than for a typical $z \sim 0.5$ lens. Large sources of additional convergence on the lens plane would require a similarly large surface mass density - we see no evidence for any suitably massive galaxies near to the lens. The low external shear in our model is also consistent with none of the group members being massive enough to generate an important amount of extra convergence.


\subsection{Cosmology}

Our model assumes the best fit cosmological parameters for the fiducial $\Lambda$CDM cosmology derived by the Planck spacecraft \cite{planckcosmo}. This cosmology assumes cold dark matter and a cosmological constant. The cosmology enters into the constraint on $\gamma$ as the critical lensing surface mass density is proportional to $D_{0,zl} D_{0,zs} / D_{zl,zs}$. The low redshift of the lens means that the angular diameter distance to the lens is only sensitive to the Hubble constant,  $H_0$. Because the source is at much higher redshift that the lens $D_{0,zs} / D_{zl,zs} \approx 1$ regardless of the cosmological parameters. Combining these two effects, our inference on the lensing mass is therefore inversely proportional to the assumed value of $H_0$. The value of the Hubble constant inferred from Planck \cite{planckcosmo} has an uncertainty of 1.3\% implying an uncertainty of 2.6\% on $\gamma$. The Planck measurements are derived assuming $\gamma=1$, it is therefore possible that if $\gamma \neq 1$ the inferred value of $H_0$ may not be the same as that derived assuming $\Lambda$CDM. In this case our inference on $\gamma$ is inversely proportional to the change in $H_0$
\be
\gamma_\mathrm{True}=\gamma_\mathrm{inferred} \frac{H_{0,\Lambda \mathrm{CDM}}}{H_{0,\mathrm{True}}}
\ee
Where $\gamma_\mathrm{True}$ is the real value of $\gamma$, $\gamma_\mathrm{inferred}$ is the value inferred assuming the $\Lambda$CDM value of $H_0$,  $H_{0,\Lambda \mathrm{CDM}}$ \cite{planckcosmo} and ${H_{0,\mathrm{True}}}$ is the correct value of the Hubble Constant.

\subsection{Uncertainties on kinematics}
\label{sec:kinsys}

As noted in Section~\ref{sec:muse}, the inference on the stellar velocity dispersions is systematically lower by $2.2\%$ when the CvD stellar templates are used compared to either the Indo-US or MILES templates. This systematic uncertainty reflects our limited knowledge of the intrinsic properties of the stars. Propagating this uncertainty onto the value of $\gamma$, we find that the $2.2\%$ uncertainty on velocity dispersion produces a $8.8\%$ uncertainty on $\gamma$.

\subsection{Final inference on $\gamma$}

After combining the statistical uncertainty from the sampling in quadrature with the 0.01 systematic modelling uncertainty, the 0.026 uncertainty from cosmology and the 0.088 uncertainty from the kinematic fit, we infer that the true inference on the ratio of the two potentials is $\gamma = 0.97 \pm 0.09$. Figure \ref{fig:pofgamma} shows the full posterior distribution $P(\gamma)$. The uncertainty on the lensing mass is 1.3 percent, dominated by uncertainty on the Hubble constant, whilst the uncertainty on the dynamical mass is 4.5 percent dominated by the uncertainty on the velocity dispersions which comes from the choice of spectral template library. The velocity dispersion uncertainty dominates the total error budget for our inference on $\gamma$.

\newpage 

\renewcommand\thefigure{S\arabic{figure}}
\setcounter{figure}{0}

\begin{figure*}
  \centering
    \includegraphics[width=\textwidth,clip=True]{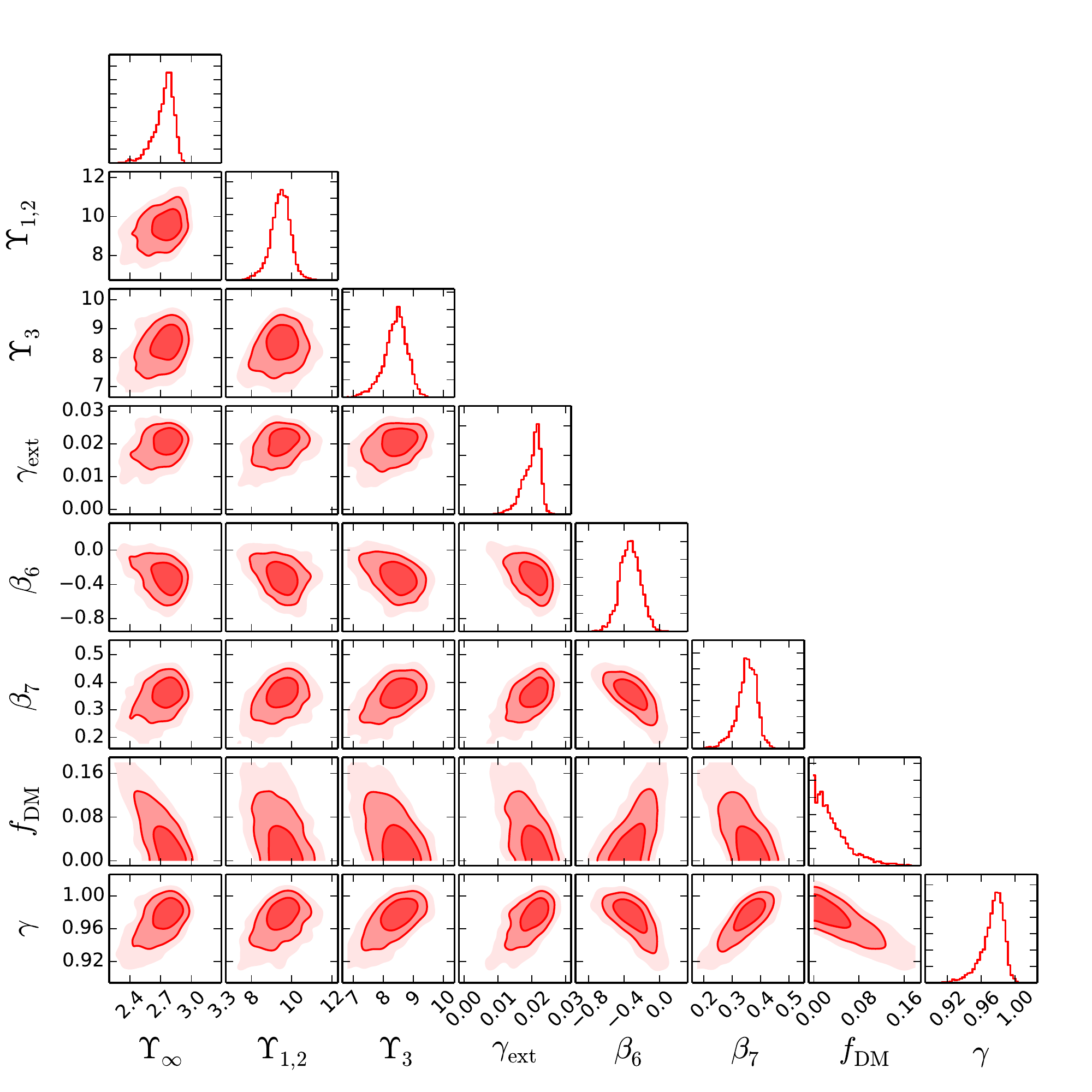}
    \caption {{\bf Two-dimensional posterior distributions for the parameters of our fiducial model for \lensp.} Only parameters from Table 2 that have a strong degeneracy with $\gamma$ are shown. Contours are the 68, 95 and 99.8 \% confidence regions. These posteriors do not include systematics from model choice or spectral template library choice for the velocity dispersion measurement. The $\Upsilon_i$ values parameterise the mass-to-light profile but individual $\Upsilon_i$ should not be interpreted as a physical mass-to-light ratio; Figure \ref{fig:MLprofile} should be used for this purpose. Similarly the $\beta_i$ values parameterise the anisotropy profile but individual $\beta_i$ are not physically meaningful quantities; Figure \ref{fig:beta} should be used for a physical interpretation of the anisotropy profile.}
    \label{fig:corner}
\end{figure*}

\begin{figure}
  \centering
    \includegraphics[width=\columnwidth,clip=True]{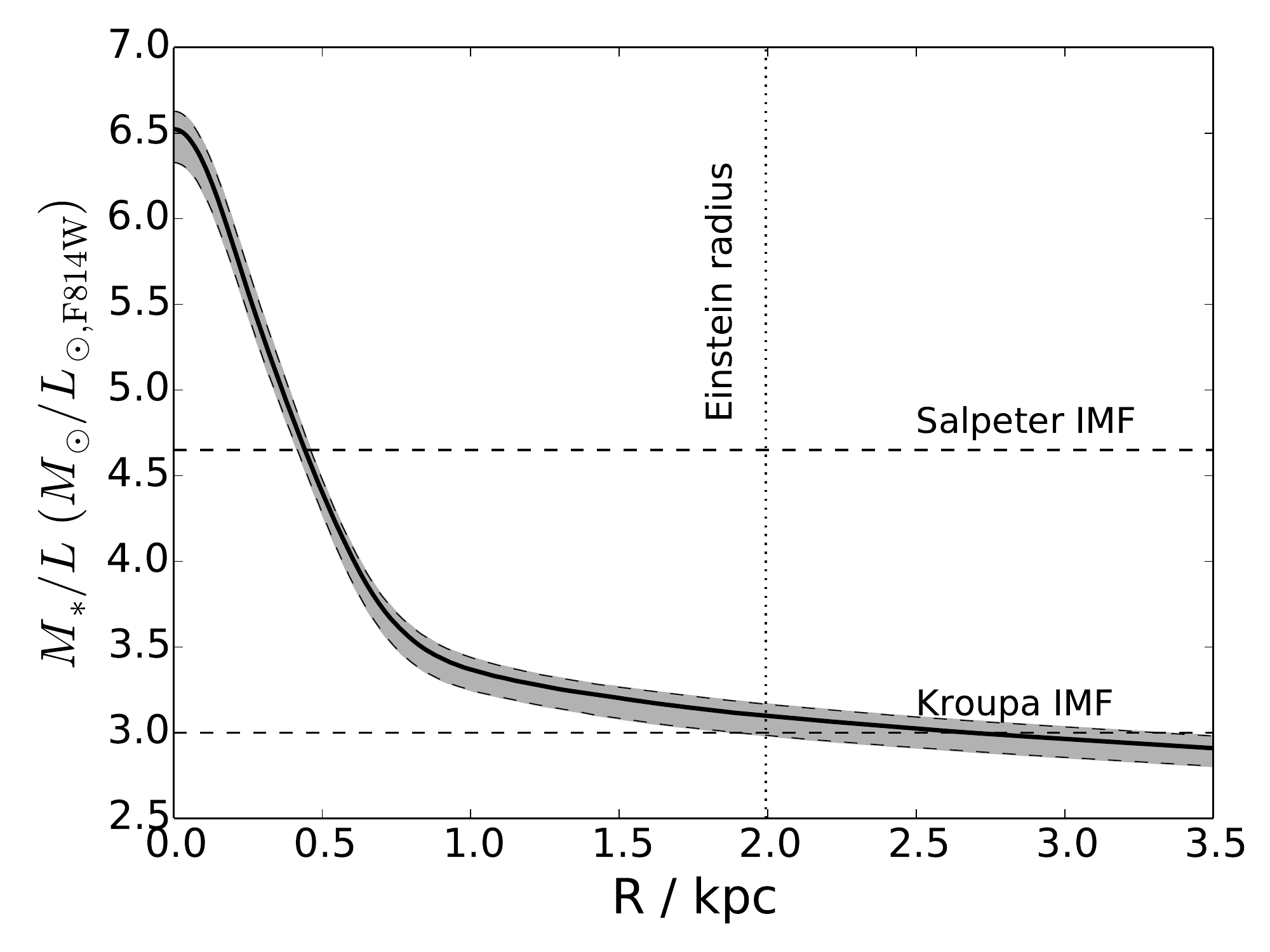}
    \caption {{\bf The projected radial stellar mass-to-light profile of \lensp} as inferred from our fiducial model, which assumes an NFW dark matter halo. The solid line shows the median of the fit model, whilst the grey band shows the 68 \% confidence region. The vertical line indicates the Einstein radius and the horizontal dashed lines indicate the expected stellar mass-to-light ratio for a Kroupa (Milky Way like) and a Salpeter (Elliptical galaxy like) stellar initial mass function.}
    \label{fig:MLprofile}
\end{figure}

\begin{figure*}
  \centering
    \includegraphics[width=\textwidth,clip=True]{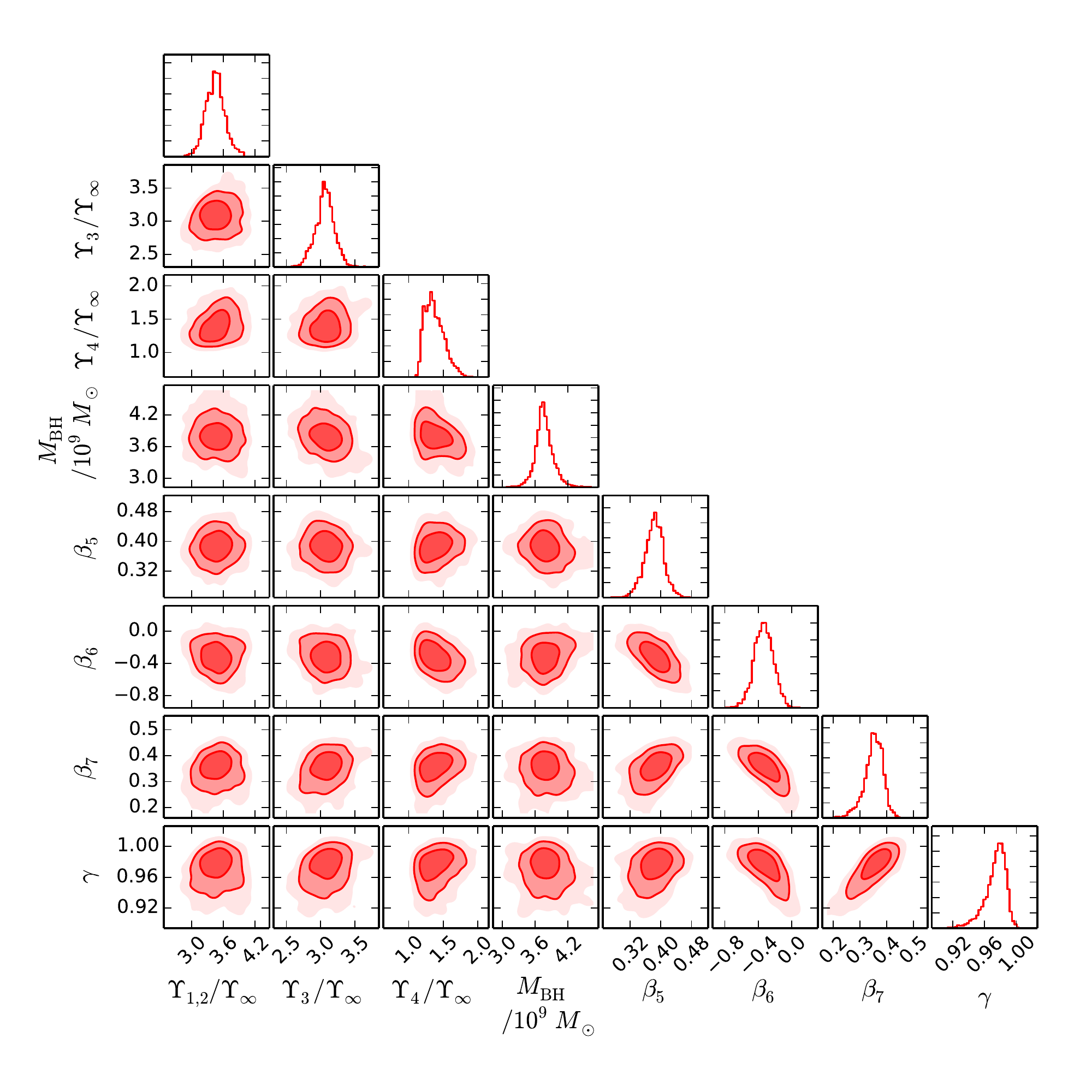}
    \caption {{ \bf Two-dimensional posteriors for the mass-to-light gradient parameters.} The mass-to-light gradient ($\Upsilon_i/\Upsilon_\infty \neq 1$) is detected at high significance and is insensitive to the GR test parameter $\gamma$. The mass-to-light gradient parameters are also not strongly degenerate with the black hole mass (shown here in units of $10^9 M_\odot$) or anisotropy profile of the lens. }
    \label{fig:cornerML}
\end{figure*}

\begin{figure}
  \centering
    \includegraphics[width=\columnwidth,clip=True]{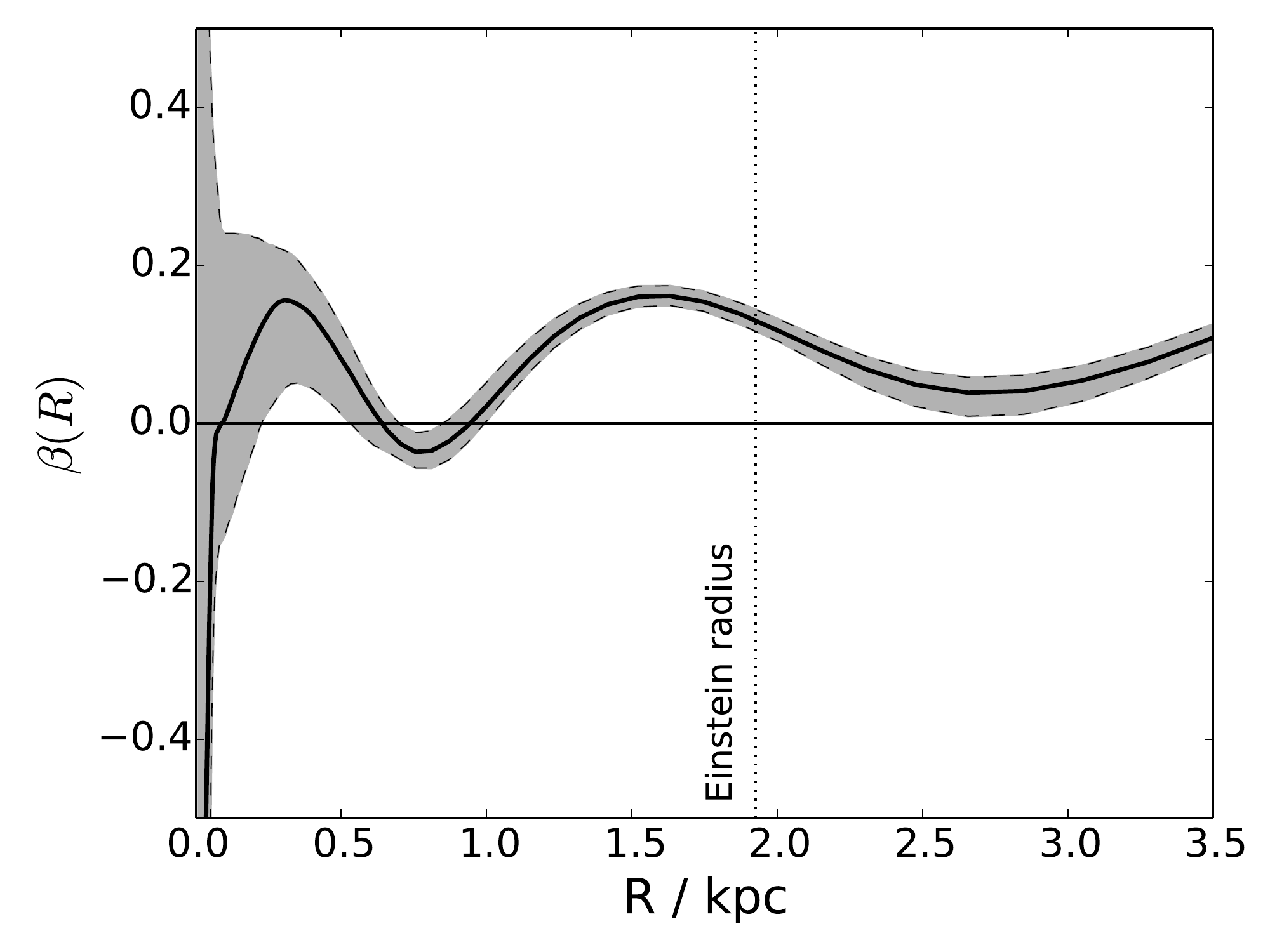}
    \caption {{\bf The inferred radial anisotropy profile of \lensp.} As in figure \ref{fig:MLprofile}, but for the orbital anisotropy of the stars. The horizontal line corresponds to isotropic orbits}
    \label{fig:beta}
\end{figure}

\begin{figure}
  \centering
    \includegraphics[width=\columnwidth,clip=True]{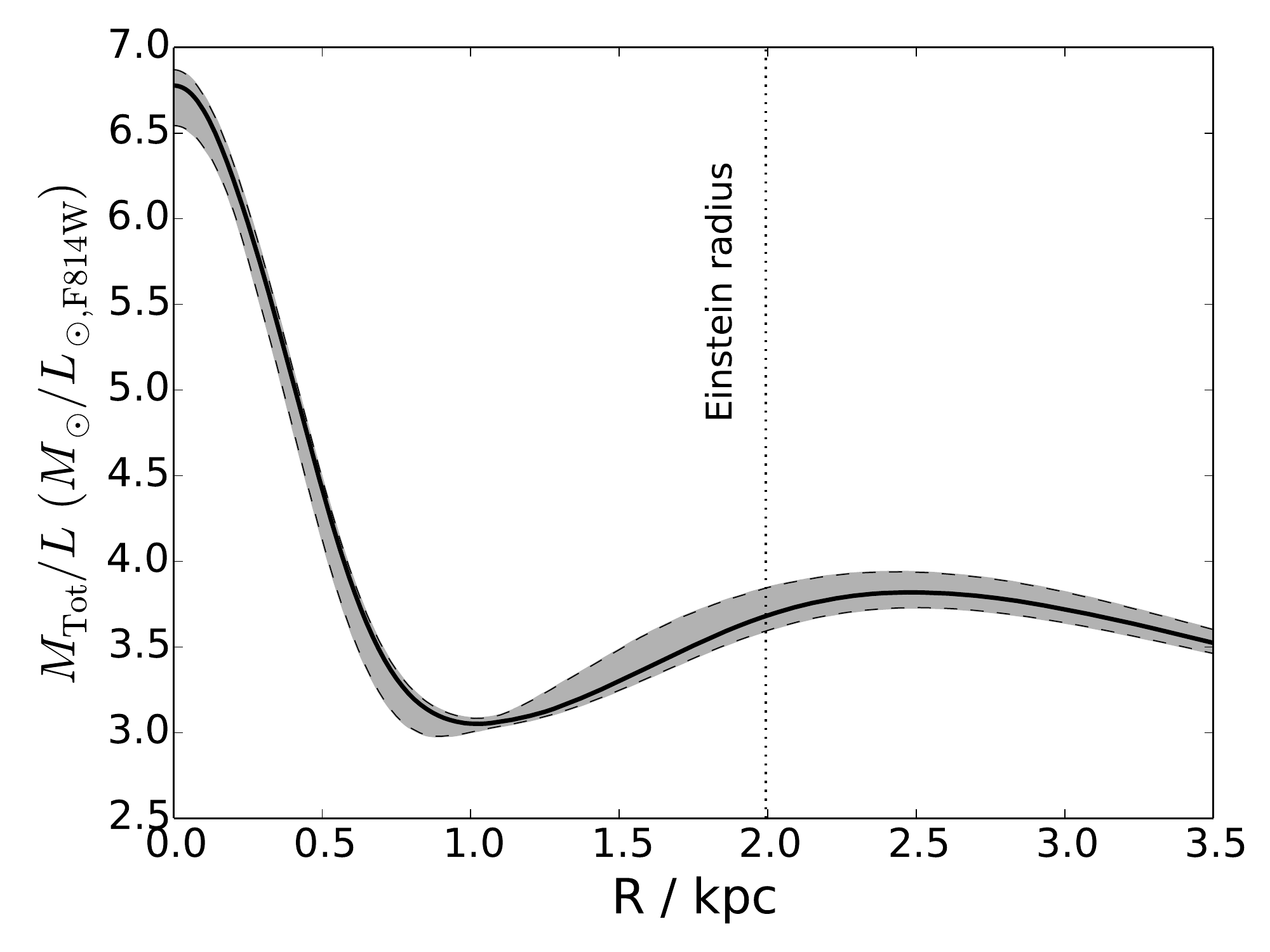}
    \caption {{\bf The projected radial total mass-to-light profile of \lensp.} As in figure \ref{fig:MLprofile}, but for the total mass-to-light profile. This is inferred from an alternative mass model without a dark matter halo and a free mass-to-light ratio for each MGE component.}
    \label{fig:remcoland}
\end{figure}

\begin{figure}
  \centering
    \includegraphics[width=\columnwidth,clip=True]{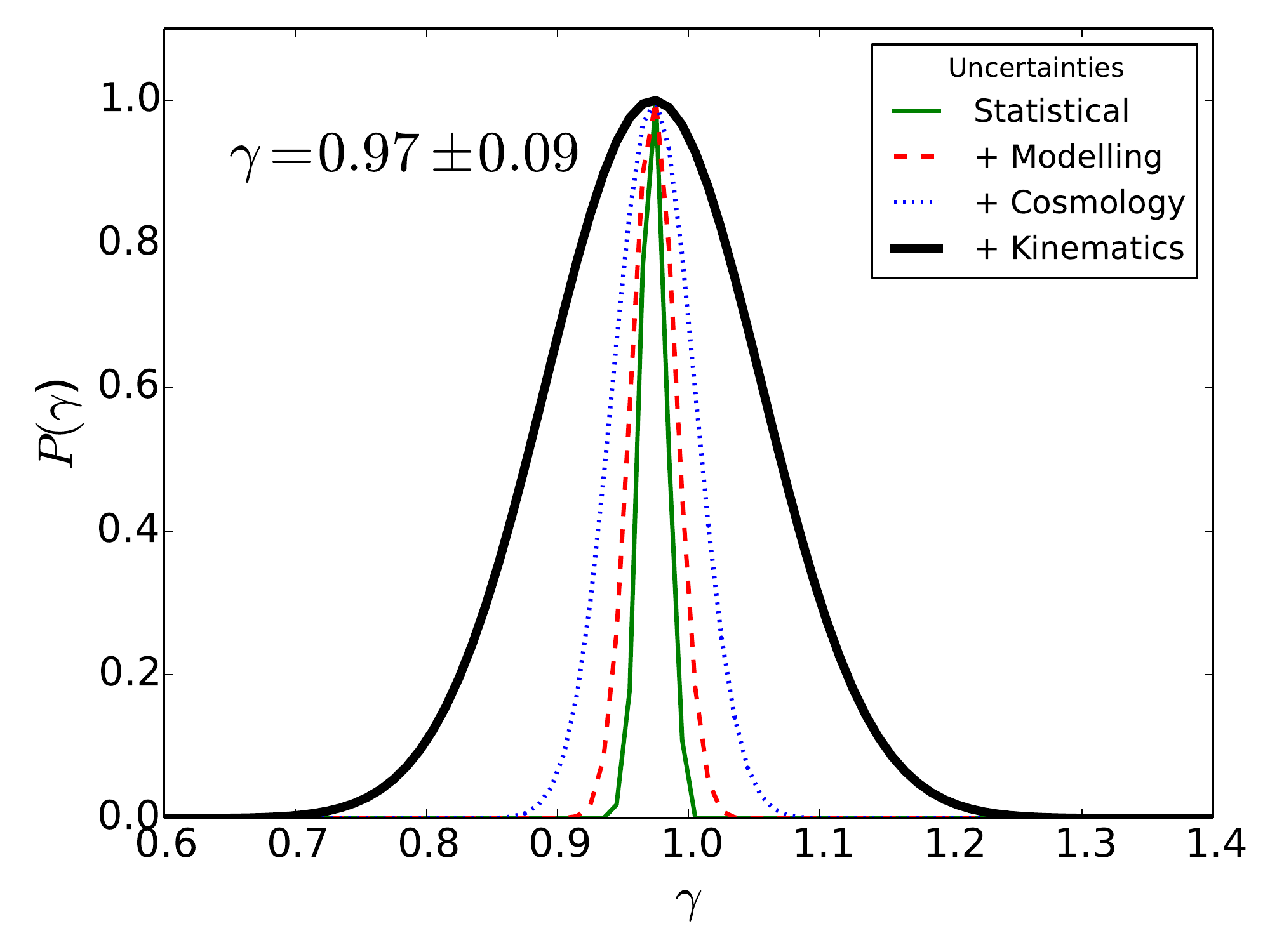}
    \caption {{\bf The relative probability density for $\gamma$ after accounting for statistical and systematic uncertainties.} The green line shows the statistical uncertainty for the fiducial model. The red line additionally includes the effect of systematics from the choice of lens model related to the assumed dark matter profile. The blue line adds the effect of uncertainty in the value of the cosmological parameters. The thick black line adds the dominant systematic: the choice of spectral library when measuring the velocity dispersion in our data.}
    \label{fig:pofgamma}
\end{figure}


\begin{thebibliography}{}

\bibitem{planckcosmo} Planck Collaboration, Ade, P.~A.~R., Aghanim, N., et al.\ Planck 2015 results. XIII. Cosmological parameters
, 2016, \aap, 594, A13 

\bibitem{Bertschinger2011} Bertschinger E.\ One 
gravitational potential or two? Forecasts and tests, 2011, Philos. Trans. R. Soc. London Ser. A, 369, 4947 


\bibitem{Hu2007} Hu W., Sawicki I.\ Models of f(R) cosmic acceleration that evade solar system tests, 2007, \prd, 76, 064004 

\bibitem{Bertotti2003} Bertotti B., Iess L., Tortora P.\ A test of general relativity using radio links with the Cassini spacecraft, 2003, \nat, 425, 374 


\bibitem{Song2011} Song Y.-S., et al.\ Complementarity 
of weak lensing and peculiar velocity measurements in testing general 
relativity, 2011, \prd, 84, 083523 


\bibitem{Simpson2013} Simpson F., et al.\ CFHTLenS: 
testing the laws of gravity with tomographic weak lensing and 
redshift-space distortions, 2013, \mnras, 429, 2249 

\bibitem{Blake2016} Blake C., et al.\ RCSLenS: testing 
gravitational physics through the cross-correlation of weak lensing and 
large-scale structure, 2016, \mnras, 456, 2806 



\bibitem{Wilcox2015} Wilcox H., et al.\ The XMM 
Cluster Survey: testing chameleon gravity using the profiles of clusters,
2015, \mnras, 452, 1171 


\bibitem{clashgr} Pizzuti L., et al.\ CLASH-VLT: 
testing the nature of gravity with galaxy cluster mass profiles, 2016, 
JCAP, 4, 023 


\bibitem{Bolton2006} Bolton A.~S., Rappaport S., Burles S.\ Constraint on the post-Newtonian parameter {$\gamma$} on galactic size scales,  2006, \prd, 74, 061501 


\bibitem{Schwab2010} Schwab J., Bolton A.~S., Rappaport S.~A.\ Galaxy-Scale Strong-Lensing Tests of Gravity and Geometric Cosmology: Constraints and Systematic Limitations, 2010, \apj, 708, 750 


\bibitem{Cao2017} Cao S., et al.\ Test of Parameterized 
Post-Newtonian Gravity with Galaxy-scale Strong Lensing Systems, 2017, \apj, 
835, 92 



\bibitem{Smith2005} Smith R.~J., Blakeslee J.~P., Lucey J.~R., Tonry J.\ Discovery of Strong Lensing by an Elliptical Galaxy at z = 0.0345, 2005, \apjl, 625, L103 





\bibitem{Smith2013} Smith R.~J., Lucey J.~R.\ A giant elliptical galaxy with a lightweight initial mass function,  2013, \mnras, 434, 1964 


\bibitem{si} Methods are available as supplementary materials on Science Online.


\bibitem{Collettthesis} Collett, T. E., Constraining Cosmology with Multiple Plane Strong Gravitational Lenses, PhD Thesis, University of Cambridge, 2015.



\bibitem{Suyu2013} Suyu, S.~H., Auger, M.~W., Hilbert, S., et al.\  Two Accurate Time-delay Distances from Strong Lensing: Implications for Cosmology, 2013, \apj, 766, 70 


\bibitem{Vegetti2010} Vegetti S., Koopmans L.~V.~E., Bolton A., Treu T., Gavazzi R.\ Detection of a dark substructure through gravitational imaging, 2010, \mnras, 408, 1969 


\bibitem{Collett2014} Collett T.~E., Auger M.~W.\ Cosmological constraints from the double source plane lens SDSSJ0946+1006,  2014, \mnras, 443, 969 

\bibitem{Wong2017} Wong K.~C., et al.\ H0LiCOW - IV. Lens mass model of HE 0435-1223 and blind measurement of its time-delay distance for cosmology,  2017, \mnras, 465, 4895 


\bibitem{Bacon2010} Bacon R., et al.\ The MUSE 
second-generation VLT instrument, Proc. SPIE, Volume 7735, id. 773508 (2010)


\bibitem{Cappellari2008} Cappellari M.\ Measuring the 
inclination and mass-to-light ratio of axisymmetric galaxies via 
anisotropic Jeans models of stellar kinematics, 2008, \mnras, 390, 71 



\bibitem{emcee} Foreman-Mackey 
D., Hogg D.~W., Lang D., Goodman J.\ emcee: The MCMC Hammer,  2013, \pasp, 
125, 306 








\bibitem{McConnell2013} McConnell N.~J., Ma C.-P.\ Revisiting the Scaling Relations of Black Hole Masses and Host Galaxy Properties,  2013, \apj, 764, 184 



\bibitem{Auger2010b} Auger M.~W., et al.\ Dark Matter 
Contraction and the Stellar Content of Massive Early-type Galaxies: 
Disfavoring ``Light'' Initial Mass Functions,  2010, \apjl, 721, L163 

\bibitem{nfw} Navarro J.~F., Frenk C.~S., White S.~D.~M.\ The Structure of Cold Dark Matter Halos,  1996, \apj, 462, 563 


\bibitem{Blumenthal1986} Blumenthal G.~R., Faber 
S.~M., Flores R., Primack J.~R.\ Contraction of dark matter galactic halos 
due to baryonic infall,  1986, \apj, 301, 27 

\bibitem{Gnedin2004} Gnedin O.~Y., Kravtsov A.~V., 
Klypin A.~A., Nagai D.\ Response of Dark Matter Halos to Condensation of 
Baryons: Cosmological Simulations and Improved Adiabatic Contraction Model,  
2004, \apj, 616, 16 

\bibitem{Abadi2010} Abadi M.~G., Navarro J.~F., Fardal 
M., Babul A., Steinmetz M.\ Galaxy-induced transformation of dark matter 
haloes,  2010, \mnras, 407, 435 




\bibitem{CvD12} Conroy C., van Dokkum P.\ Counting Low-mass Stars in Integrated Light,  2012, \apj, 747, 69 

\bibitem{INDOUS} Valdes F., Gupta R., Rose J.~A., 
Singh H.~P., Bell D.~J.\ The Indo-US Library of Coud{\'e} Feed Stellar 
Spectra,  2004, \apjs, 152, 251 






\bibitem{esorex}  https://www.eso.org/sci/software/cpl/esorex.html

\bibitem{Oldham2017b} Oldham L., et al.\ The fundamental plane of evolving red nuggets,  2017, \mnras, 470, 3497 

\bibitem{MILES} S{\'a}nchez-Bl{\'a}zquez, P., Peletier, R.~F., Jim{\'e}nez-Vicente, J., et al.\ Medium-resolution Isaac Newton Telescope library of empirical spectra, 2006, \mnras, 371, 703 


\bibitem{Newman2017} Newman A.~B., Smith R.~J., Conroy 
C., Villaume A., van Dokkum P.\ The Initial Mass Function in the Nearest 
Strong Lenses from SNELLS: Assessing the Consistency of Lensing, Dynamical, 
and Spectroscopic Constraints,  2016, arXiv:1612.00065 


\bibitem{Greene2006} Greene J.~E., Ho L.~C.\ Measuring Stellar Velocity Dispersions in Active Galaxies,  2006, \apj, 641, 117 


\bibitem{GO10429} Hubble Space Telescope General Observer Programme 10429, PI: Blakeslee.

\bibitem{GO10710} Hubble Space Telescope General Observer Programme 10710, PI: Noll.


\bibitem{Fruchter2002} Fruchter A.~S., Hook R.~N.\ Drizzle: A Method for the Linear Reconstruction of Undersampled Images,  2002, \pasp, 114, 144 


\bibitem{drizzle} Jedrzejewski R., Hack 
W., Hanley C., Busko I., Koekemoer A.~M.\ MultiDrizzle: Automated Image 
Combination and Cosmic-Ray Identification Software,  2005, Astronomical 
Data Analysis Software and Systems XIV, 347, 129 


\bibitem{Emsellem1994} Emsellem E., Monnet G., Bacon R.\ The multi-gaussian expansion method: a tool for building realistic photometric and kinematical models of stellar systems I. The formalism,  1994, \aap, 285, 723 

\bibitem{Cappellari2002} Cappellari M.\ Efficient 
multi-Gaussian expansion of galaxies,  2002, \mnras, 333, 400 



\bibitem{Barnabe2012} Barnab{\`e} M., et al.\ The 
SWELLS survey - IV. Precision measurements of the stellar and dark matter 
distributions in a spiral lens galaxy,  2012, \mnras, 423, 1073 

\bibitem{Munoz2001} Mu{\~n}oz J.~A., Kochanek C.~S., Keeton C.~R.\ Cusped Mass Models of Gravitational Lenses,  2001, \apj, 558, 657 



\bibitem{Barkana1998} Barkana R.\ Fast Calculation of a 
Family of Elliptical Mass Gravitational Lens Models,  1998, \apj, 502, 531 

\bibitem{Kravtsov2013} Kravtsov A.~V.\ The Size-Virial 
Radius Relation of Galaxies,  2013, \apjl, 764, L31 

\bibitem{Warren2003} Warren S.~J., Dye S.\ Semilinear Gravitational Lens Inversion,  2003, \apj, 590, 673 


\bibitem{Suyu2006} Suyu S.~H., Marshall P.~J., Hobson M.~P., Blandford R.~D.\ A Bayesian analysis of regularized source inversions in gravitational lensing, 2006, \mnras, 371, 983 


\bibitem{Cappellari2012} Cappellari, M., McDermid, R.~M., Alatalo, K., et al.\ Systematic variation of the stellar initial mass function in early-type galaxies, 2012, Nature, 484, 485 

\bibitem{LaBarbera2015} La Barbera F., Ferreras I., Vazdekis A.\ The initial mass function of early-type galaxies: no correlation with [Mg/Fe],  2015, \mnras, 449, L137 


\bibitem{MartinNavarro2015} Mart{\'{\i}}n-Navarro I., Barbera F.~L., Vazdekis A., Falc{\'o}n-Barroso J., Ferreras I.\ Radial variations in the stellar initial mass function of early-type galaxies,  2015, \mnras, 447, 1033 

\bibitem{vdc2017} van Dokkum, P., Conroy, C., Villaume, A., Brodie, J., \& Romanowsky, A.~J.,\ The Stellar Initial Mass Function in Early-type Galaxies from Absorption Line Spectroscopy. III. Radial Gradients, 2017, \apj, 841, 68 



\bibitem{Conroy2012b} Conroy C., van Dokkum P.~G.\ The Stellar Initial Mass Function in Early-type Galaxies From Absorption Line Spectroscopy. II. Results,  2012, \apj, 760, 71 




\bibitem{OldhamM87} Oldham L.~J., Auger M.~W.\ Galaxy structure from multiple tracers - II. M87 from parsec to megaparsec scales,  2016, \mnras, 457, 421 


\bibitem{Auger2010a} Auger M.~W., et al.\ The Sloan Lens 
ACS Survey. X. Stellar, Dynamical, and Total Mass Correlations of Massive 
Early-type Galaxies,  2010, \apj, 724, 511 


\bibitem{Cappellari2007} Cappellari, M., Emsellem, E., Bacon, R., et al.\ The SAURON project - X. The orbital anisotropy of elliptical and lenticular galaxies: revisiting the (V/$\sigma$, $\epsilon$) diagram with integral-field stellar kinematics, 2007, \mnras, 379, 418 


\bibitem{Wyithe2001} Wyithe J.~S.~B., Turner E.~L., Spergel D.~N.\ Gravitational Lens Statistics for Generalized NFW Profiles: Parameter Degeneracy and Implications for Self-Interacting Cold Dark Matter,  2001, \apj, 555, 504 


\bibitem{Duffy2010} Duffy A.~R., et al.\ Impact of 
baryon physics on dark matter structures: a detailed simulation study of 
halo density profiles,  2010, \mnras, 405, 2161 



\bibitem{MSD} Falco E.~E., Gorenstein M.~V., Shapiro I.~I.\ On model-dependent bounds on H(0) from gravitational images Application of Q0957 + 561A,B,  1985, \apjl, 289, L1 

\bibitem{Wong2011} Wong K.~C., Keeton C.~R., Williams 
K.~A., Momcheva I.~G., Zabludoff A.~I.\ The Effect of Environment on Shear 
in Strong Gravitational Lenses,  2011, \apj, 726, 84 


\bibitem{Collett2013} Collett T.~E., et al.\ 
Reconstructing the lensing mass in the Universe from photometric catalogue 
data 2013, \mnras, 432, 679 

\bibitem{Keeton2003} Keeton, C.~R.\ Analytic Cross Sections for Substructure Lensing, 2003, \apj, 584, 664 



\bibitem{Smith2015} Smith R.~J., Lucey J.~R., Conroy C.\ The SINFONI Nearby Elliptical Lens Locator Survey: discovery of two new low-redshift strong lenses and implications for the initial mass function in giant early-type galaxies, 2015, \mnras, 449, 3441 

\bibitem{Wu2014} Wu, X., Gerhard, O., Naab, T., et al.\ The mass and angular momentum distribution of simulated massive early-type galaxies to large radii, 2014, \mnras, 438, 2701 


\bibitem{Rusu2017} Rusu, C.~E., Fassnacht, C.~D., Sluse, D., et al.\  H0LiCOW - III. Quantifying the effect of mass along the line of sight to the gravitational lens HE 0435-1223 through weighted galaxy counts, 2017, \mnras, 467, 4220 






\end{thebibliography}
\end{document}